\documentclass[11pt, draftclsnofoot, onecolumn]{IEEEtran}
\usepackage{setspace}
\usepackage{caption}
\usepackage{graphicx}
\usepackage{array}
\usepackage{booktabs}
\usepackage{caption}
\usepackage{ragged2e}
\usepackage{tabularx}
\usepackage{epstopdf}
\usepackage{amsmath, amsthm, amssymb, amsfonts, mathtools}
\usepackage{cases, multirow}
\usepackage{cite}
\usepackage{courier}
\usepackage{subfigure}
\usepackage{float}
\usepackage{color}
\usepackage{multicol}
\usepackage{dblfloatfix}
\usepackage{cuted, nccmath}
\usepackage{lipsum}
\usepackage{mathtools}
\usepackage{threeparttable}
\long\def\comment#1{}

\newtheorem{thm}{Theorem}

\def\figref#1{Fig.~\ref{#1}}
\def\be{\begin{equation} }
\def\ee{\end{equation}}

\title{On Shadowing the $\kappa$-$\mu$ Fading Model}
\begin{document}

%\author{The Authors
%}

\author{
Nidhi Simmons,~\IEEEmembership{Member,~IEEE}, Carlos Rafael Nogueira da Silva, Simon~L.~Cotton,~\IEEEmembership{Senior Member,~IEEE}, Paschalis~C.~Sofotasios,~\IEEEmembership{Senior Member,~IEEE}, Seong Ki Yoo,~\IEEEmembership{Member,~IEEE} and Michel Daoud Yacoub,~\IEEEmembership{Member,~IEEE}

%\IEEEcompsocitemizethanks{
  %\IEEEcompsocthanksitem 
	%%This work was supported by the U.K. Engineering and Physical Sciences Research Council under Grant Reference EP/L026074/1, the Department for the Economy Northern Ireland under Grant USI080 and by CNPq under Grant Reference 304248/2014-2.
	%
  %N. Bhargav and S. L. Cotton are with the Wireless Communications Laboratory, Institute of Electronics, Communications and Information Technology, The Queen's University of Belfast, Queen's Road, Belfast, BT3 9DT, UK. E-mail: \{nbhargav01, simon.cotton\}@qub.ac.uk.
	%
  %Carlos Rafael Nogueira da Silva and Michel Daoud Yacoub are with the Wireless Technology Laboratory, School of Electrical
%and Computer Engineering, University of Campinas, Campinas 13083-970,
%Brazil. E-mail: \{carlosrn, michel\}@decom.fee.unicamp.br.
	%
	%Paschalis C. Sofotasios is with the Department of Electrical and Computer Engineering, Khalifa University of Science and Technology, PO Box 127788, Abu Dhabi, UAE and also with the Department of Electronics and Communications Engineering, Tampere University of Technology, 33101 Tampere, Finland. E-mail: \{p.sofotasios\}@ieee.org. 
	%
	%This work has been submitted to the IEEE for possible publication. Copyright may be transferred without notice, after
%which this version may no longer be accessible.
%
  %}
}

\maketitle
\begin{abstract}
In this paper, we extensively investigate the way in which $\kappa$-$\mu$ fading channels can be impacted by shadowing. Following from this, a family of shadowed $\kappa$-$\mu$ fading models are introduced and classified according to whether the underlying $\kappa$-$\mu$ fading undergoes  single or double shadowing.
In total, we discuss three types of \textit{single} shadowed $\kappa$-$\mu$ model (denoted Type~I to Type~III) and three types of \textit{double} shadowed $\kappa$-$\mu$ model (denoted Type~I to Type~III).
%Many of these shadowed fading models are introduced in the literature for the first time.
The taxonomy of the \textit{single} shadowed Type~I~-~III models is dependent upon whether the fading model assumes that the dominant component, the scattered waves, or both experience shadowing. Although the physical definition of the examined models make no predetermination of the statistics of the shadowing process, for illustrative purposes, two example cases are provided for each type of \textit{single} shadowed model by assuming that the shadowing is influenced by either a Nakagami-$m$ random variable (RV) or an inverse Nakagami-$m$ RV.
%{\color{red} Two example cases (A and B) are discussed for each type where it is assumed that the shadowing is influenced by either a Nakagami-$m$ RV or an inverse Nakagami-$m$ RV. Analytical formulations are derived for the probability density function (PDF) of the Type~I.A and Type~II.A models, whilst a closed-form expression for the PDF of the Type~II.B model is obtained.}  %
The categorization of the~\textit{double} shadowed Type I - III models is dependent upon whether a) the envelope experiences shadowing of the dominant component, which is preceded (or succeeded) by a secondary round of shadowing (multiplicative), or b) the dominant and scattered contributions are fluctuated by two independent shadowing processes, or c) the scattered waves of the envelope are subject to shadowing, which is also preceded (or succeeded) by a secondary round of multiplicative shadowing. Similar to the \textit{single} shadowed models, we provide two example cases for each type of \textit{double} shadowed model by assuming that the shadowing phenomena are shaped by a Nakagami-$m$ RV, an inverse Nakagami-$m$ RV or their mixture.   
It is worth highlighting that the \textit{double} shadowed $\kappa$-$\mu$ models offer remarkable flexibility as they include the $\kappa$-$\mu$, $\eta$-$\mu$, and the various types of \textit{single} shadowed $\kappa$-$\mu$ distribution as special cases. Moreover, we demonstrate a practical application of the \textit{double} shadowed $\kappa$-$\mu$ Type~I model by applying it to realistic channel measurements obtained for body area networks operating at 2.45 GHz.

\end{abstract}

\begin{IEEEkeywords}
BAN, channel modeling, generalized fading, shadowed $\kappa$-$\mu$ fading, shadowing.
\end{IEEEkeywords}

%%%%%%%%%%%%%%%%%%%%%%%%%%%%%%%%%%%%%%%%%%%%%%%%%%%%%%%%%%%
\section{Introduction}
%%%%%%%%%%%%%%%%%%%%%%%%%%%%%%%%%%%%%%%%%%%%%%%%%%%%%%%%%%%
The $\kappa$-$\mu$ fading model~\cite{4231253} is a generalized fading model which was developed to describe envelope fluctuations that arise due to the clustering of scattered multipath waves in addition to the presence of elective dominant components. It is characterized by two physical fading parameters, namely $\kappa$ and $\mu$. Here, $\kappa$ represents the ratio of the total power of the dominant component to the total power of the scattered waves whilst $\mu$ represents the number of multipath clusters. Due to its inherent versatility and general nature, it contains other well-known fading models such as the Rice ($\kappa$ = $k$, $\mu$ = 1), Nakagami-$m$ ($\kappa$ $\rightarrow$ 0, $\mu$ = $m$), Rayleigh ($\kappa$ $\rightarrow$ 0, $\mu$ = 1) and One-Sided Gaussian ($\kappa$ $\rightarrow$ 0, $\mu$ = 0.5) as special cases. 

%A $\kappa$-$\mu$ fading envelope can be affected by shadowing in many different ways.

As well known, in a $\kappa$-$\mu$ fading environment, the signal reaching the receiver may contain dominant as well 
as clusters of scattered waves. These waves arise by means 
of different propagation mechanisms and may be shadowed 
by common or distinct obstacles.
For instance, the dominant component, the scattered waves, or both can be impacted by this propagation phenomenon.  It is also entirely possible that in addition to the dominant component being shadowed, further multiplicative shadowing\footnote{In this case, the total power of the dominant and scattered signal components are shadowed.} may occur which impacts the scattered signal, and also administers secondary shadowing to the already perturbed dominant component. Likewise, in addition to the scattered waves being shadowed, further shadowing may occur which impacts the dominant signal component, and administers secondary shadowing to the fluctuated scattered waves. As well as this, both the dominant component and scattered waves can be influenced by individual shadowing processes. Hence, a number of 
shadowing combinations give rise to a family of shadowed 
$\kappa$-$\mu$ fading models that can be classified depending on whether the underlying $\kappa$-$\mu$ fading undergoes single or double shadowing. Motivated by this, the aim of this paper is to explore the different ways of shadowing 
the $\kappa$-$\mu$ fading model.
%These different ways in which shadowing can impact the envelope fading leads to a family of shadowed $\kappa$-$\mu$ fading models that can be classified depending on whether the underlying $\kappa$-$\mu$ fading undergoes single or double shadowing. 

Traditionally, shadowing has been modeled using the lognormal distribution~\cite{1175472}. However, due to challenges which exist in relation to its tractability, the authors in~\cite{abdi1999utility} proposed the use of the gamma distribution. Similarly,~\cite{6594884, cotton2015human} proposed the use of the closely related Nakagami-$m$ distribution due to its ability to exhibit semi-heavy tailed characteristics~\cite{cotton2015human}. More recently,~\cite{yoo2017fisher} and~\cite{newTCOM} used the inverse Nakagami-$m$ and inverse gamma distributions, respectively. It should be noted that similar to the lognormal, gamma and Nakagami-$m$ distributions, the inverse gamma and inverse Nakagami-$m$ distributions can also exhibit the necessary semi heavy-tailed behavior to accurately characterize shadowing. Moreover, they offer much of the analytical tractability available from using the gamma and Nakagami-$m$ distributions.

\begin{table*} 
\renewcommand{\arraystretch}{1.2}
\centering
{\caption{Physical interpretation of the single shadowed $\kappa$-$\mu$ fading models}}
\label{Table:1.0}
\begin{threeparttable}
% Preview source code for paragraph 0
\small
\begin{tabular}{|>{\centering}p{2.1cm}|>{\centering}p{2.6cm}|p{9cm}|}
\hline 
Fading models & Shadowing type &~~~~~~~~~~~~~~~~~~~~~~~~~~Physical interpretation\tabularnewline
\hline 
\hline 
Single shadowed $\kappa$-$\mu$ Type I & Shadowing of the dominant component & Physically, this situation may arise when the signal power delivered
through the optical path between the transmitter and receiver is shadowed
by objects moving within its locality. For example, blockages to the
dominant path caused by cars, buildings and/or people present/moving
within the locality of the transmitter or receiver.\tabularnewline
\hline 
Single shadowed $\kappa$-$\mu$ Type II & Shadowing of the scattered components  & Physically, this situation may arise when the scattered signal components
between the transmitter and receiver are shadowed by objects moving
within their locality. For example, blockages to the scattered components
due to cars, building and/or people present/moving within the locality
of the transmitter or receiver.\tabularnewline
\hline 
Single shadowed $\kappa$-$\mu$ Type III & Shadowing of both the dominant component and scattered waves & Physically, this situation may arise when the dominant component and
scattered signal components between the transmitter and receiver undergo
shadowing caused by objects moving within the locality of the transmitter
or receiver and/or large-scale effects.\tabularnewline
\hline 
\end{tabular}\end{threeparttable}
\end{table*}

In this work, we discuss three types of single shadowed $\kappa$-$\mu$ fading model (denoted I through to III) which assume that the multipath fading is manifested by the propagation mechanisms associated with $\kappa$-$\mu$ fading. In addition, these models consider that either the dominant component (Type I), the scattered waves (Type II), or both (Type III) suffer from a single shadowing process.
We emphasize that these model frameworks are general and make no predetermination on the random variable (RV) that is responsible for characterizing the shadowing phenomena. For illustrative purposes, we provide two example cases for each type of single shadowed $\kappa$-$\mu$ fading model where it is assumed that the shadowing is influenced by either a Nakagami-$m$ RV or an inverse Nakagami-$m$ RV. 
%which is shaped by a Nakagami-$m$ or an inverse Nakagami-$m$ random variable (RV). 
%Note that the single shadowed $\kappa$-$\mu$ Type~I.B (where the dominant component is shadowed by an inverse Nakagami-$m$ RV), Type~II.A~and~B (where the scattered waves are shadowed by a Nakagami-$m$ RV and an inverse Nakagami-$m$ RV, respectively) models are presented here for the first time. 
We also introduce three types of double shadowed $\kappa$-$\mu$ fading model, denoted I through to III. 
The Type~I model in this case assumes that in addition to the dominant component of a $\kappa$-$\mu$ signal being shadowed, further shadowing also occurs which impacts the scattered signal and also administers secondary shadowing to the already perturbed dominant component. 
Therefore, this model provides a convenient way to not only control the shadowing of the dominant component, but also any multiplicative shadowing\footnote{In this case, the total power of the dominant and scattered signal components are shadowed.} which may be present in practical wireless channels. 
The Type~II model considers that the dominant component and scattered waves of a $\kappa$-$\mu$ fading envelope are perturbed by two different shadowing processes.
Lastly, the Type~III model assumes that in addition to the scattered waves of a $\kappa$-$\mu$ signal being shadowed, the rms power of the dominant component and scattered waves also experience a secondary round of shadowing. Similar to the single shadowed models, two example cases for each of the three types of double shadowed model are discussed where it is assumed that the shadowing is shaped by either a Nakagami-$m$ RV, an inverse Nakagami-$m$ RV or their mixture.  For the reader's convenience, Tables~I and~II summarize the various types of single shadowed and double shadowed $\kappa$-$\mu$ models introduced in this paper.
\begin{table*} 
\small
\renewcommand{\arraystretch}{1.2}
\centering
{\caption{Physical interpretation of the double shadowed $\kappa$-$\mu$ fading models}}
\label{Table:1.0}
\begin{threeparttable}
\begin{tabular}{|>{\centering}p{2.1cm}|>{\centering}p{2.6cm}|p{9cm}|}
\hline 
Fading models & Shadowing type &~~~~~~~~~~~~~~~~~~~~~~~~~~Physical interpretation\tabularnewline
\hline 
\hline 
Double shadowed $\kappa$-$\mu$ Type I & Shadowing of the dominant component and secondary round of multiplicative
shadowing & Physically, this situation may arise when the signal power delivered
through the optical path between the transmitter and receiver is shadowed
by objects moving within its locality (e.g., blockages to the dominant
component due to cars, buildings and/or people), whilst further shadowing
of the received power (combined multipath and dominant paths) may
also occur due to obstacles moving in the vicinity of the transmitter
or receiver and/or large-scale effects.\tabularnewline
\hline 
Double shadowed $\kappa$-$\mu$ Type II & Independent shadowing of the dominant component and scattered waves & Physically, this situation may arise when the dominant component and
scattered signal components between the transmitter and receiver undergo
independent shadowing caused by objects moving within their locality. \tabularnewline
\hline 
Double shadowed $\kappa$-$\mu$ Type III & Shadowing of the scattered components and secondary round of multiplicative
shadowing & Physically, this situation may arise when the scattered signal components
between the transmitter and receiver are shadowed by objects moving
within their locality. Furthermore, additional shadowing of the received
power (combined multipath and dominant paths) may occur due to obstacles
in the vicinity of the transmitter or receiver and/or large-scale
effects.\tabularnewline
\hline 
\end{tabular}
\end{threeparttable}
\end{table*}

It is worth highlighting that, due to the generality of the analysis presented here and under particular shadowing conditions, a number of the existing composite fading models found in the literature occur as special cases. 
For example, multiplicative composite fading models such as the $\kappa$-$\mu$/inverse gamma and $\eta$-$\mu$/inverse gamma fading models~\cite{newTCOM}, which assume that a $\kappa$-$\mu$ or an $\eta$-$\mu$ RV is responsible for generating the multipath fading, and  an inverse gamma RV for shaping the shadowing. In particular~\cite{newTCOM} obtained closed form expressions for the PDFs of the $\kappa$-$\mu$/inverse gamma and $\eta$-$\mu$/inverse gamma fading models. The utility of these composite fading models was also demonstrated through a series of channel measurements obtained for wearable, cellular and vehicular communications.
Likewise, some line-of-sight (LOS) composite models\footnote{Many of the models presented in the literature for which the dominant signal component is subject to shadowing are often referred to as LOS composite fading models.} 
%However, in the strict sense, they are not true composite models because the shadowing is not applied multiplicatively to all constituent parts of the fading envelope.
such as the $\kappa$-$\mu$ shadowed~\cite{6594884},~\cite{cotton2015human}\footnote{It is noted that the $\kappa$-$\mu$ shadowed fading model presented in~\cite{6594884} and \cite{cotton2015human} is a type of single shadowed $\kappa$-$\mu$ model.} and shadowed Rician~\cite{paris2010closed, abdi2003new} fading models are also found through the analysis conducted here.  The $\kappa$-$\mu$ shadowed fading model presented in~\cite{6594884} and \cite{cotton2015human} assumes that the multipath fading is due to fluctuations brought about by a $\kappa$-$\mu$ RV, whilst the dominant signal component is fluctuated by a Nakagami-$m$ RV. Notably, it includes the  $\kappa$-$\mu$, $\eta$-$\mu$ and shadowed Rician fading models as special cases. This model has been shown to provide excellent agreement with field measurements obtained for body-centric fading channels~\cite{cotton2015human}, land-mobile satellite channels~\cite{abdi2003new} and underwater acoustic channels~\cite{ruiz2011ricean}. 

The main contributions of this paper are now summarized as follows: 
\begin{itemize}
\item Firstly, we perform a broad investigation of the way in which $\kappa$-$\mu$ fading can be affected by shadowing. Subsequently, we introduce a family of shadowed $\kappa$-$\mu$ models that are classified as either single or double shadowed models. Three types of single shadowed $\kappa$-$\mu$ fading model (Type~I~-~III) and three types of double shadowed $\kappa$-$\mu$ fading model (Type I - III) are discussed.
\item Secondly, a thorough physical interpretation for all three types of single and double shadowed $\kappa$-$\mu$ models is provided.
\item Thirdly, we discuss two example cases for each type of single and double shadowed $\kappa$-$\mu$ fading model by assuming that the incurred shadowing is caused by a Nakagami-$m$ RV, an inverse Nakagami-$m$ RV or their mixture. It is worth remarking that the model frameworks discussed in this paper are general and make no presumption on the RV that is responsible for shaping the shadowing characteristics. The example cases discussed here are for illustrative purposes only.
\item Fourthly, the generality of the double shadowed $\kappa$-$\mu$ fading models are highlighted through reduction to a number of well-known special cases. In particular, these fading models unify the $\kappa$-$\mu$, $\eta$-$\mu$ and the various types of single shadowed $\kappa$-$\mu$ model. 
\item Finally, we provide an example of a practical application of the double shadowed $\kappa$-$\mu$ Type~I model by applying it to body area network (BAN) channel measurements obtained at 2.45~GHz.
\end{itemize}

The remainder of this paper is organized as follows. Section~II and III describe and formulate the various types of single and double shadowed $\kappa$-$\mu$ model, respectively. Section~IV presents some special cases of the double shadowed $\kappa$-$\mu$ models. Section V provides some numerical results and also demonstrates the utility of the double shadowed $\kappa$-$\mu$ Type~I fading model for characterizing the shadowing encountered in BAN communications channels. Lastly, some interesting concluding remarks are provided in Section~VI.

%%%%%%%%%%%%%%%%%%%%%%%%%%%%%%%%%%%%%%%%%%%%%%%%%%%%%%%%%%%
\section{Single Shadowed $\kappa$-$\mu$ Models}
%%%%%%%%%%%%%%%%%%%%%%%%%%%%%%%%%%%%%%%%%%%%%%%%%%%%%%%%%%%

In this section, we investigate a number of different ways in which the $\kappa$-$\mu$ fading envelope can be impacted by a single shadowing process. This leads to three types of single shadowed fading model, denoted Type~I to Type~III, with their physical interpretation provided in Table~I.

%%%%%%%%%%%%%%%%%%%%%%%%%%%%%%%%%%%%%%%%%%%%%%%%%%%%%%%%%%%
\subsection{Single Shadowed $\kappa$-$\mu$ Type~I Model}
%%%%%%%%%%%%%%%%%%%%%%%%%%%%%%%%%%%%%%%%%%%%%%%%%%%%%%%%%%%
Similar to the~$\kappa$-$\mu$~fading model, the single shadowed $\kappa$-$\mu$ Type~I fading model assumes that the received signals are composed of clusters of multipath waves propagating in non-homogeneous environments. Within each multipath cluster, the scattered waves have similar delay times and the delay spreads of different clusters are relatively large. The power of the scattered waves in each cluster is assumed to be identical whilst the power of the dominant component is assumed to be arbitrary. Unlike the $\kappa$-$\mu$ model, the single shadowed $\kappa$-$\mu$ Type~I model assumes that the dominant component of each cluster can~randomly fluctuate because of shadowing. Its signal envelope, $R$, can be expressed in terms of the in-phase and quadrature phase components as 

\begin{equation}
R^{2}=\sum_{i=1}^{\mu}\left(X_{i}+\xi p_{i}\right)^{2}+\left(Y_{i}+\xi q_{i}\right)^{2}
\label{eq_section2_01}
\end{equation}
where $\xi$ represents a RV which is responsible for introducing the shadowing, $\mu$ is a real-valued extension related to the number of multipath clusters, $X_i$ and $Y_i$ are mutually independent Gaussian random processes with mean $\mathbb{E}\left[X_{i}\right] = \mathbb{E}\left[Y_{i}\right] = 0$ and variance $\mathbb{E}\left[ {{X_i}^2} \right] = \mathbb{E}\left[ {{Y_i}^2} \right] = {\sigma ^2}$, where $\mathbb{E}[\cdot]$ denotes the statistical expectation. Also, $p_i$ and $q_i$ are the mean values of the in-phase and quadrature phase~components of the multipath cluster $i$. 
We now consider two example cases  for the single shadowed Type I model, the details of which are discussed next.

\setlength\parindent{35pt}{\textit{Example 1)}}
 In our first example of the single shadowed $\kappa$-$\mu$ Type~I model, we assume that the dominant component of a $\kappa$-$\mu$ signal undergoes variations induced by a Nakagami-$m$ RV. Thus, in \eqref{eq_section2_01} $\xi$ represents a Nakagami-$m$ RV with shape parameter\footnote{To assist with the understanding of the models presented here, throughout the manuscript we denote $m_d$, $m_s$ and $m_t$ as the shadowing parameters which are responsible for fluctuating the dominant, scattered or total (i.e. the combined dominant and scattered) components respectively.} $m_d$ and $\mathbb{E}\left[{\xi}^2 \right] = 1$.
It is worth highlighting that this model was introduced as a generalization of the $\kappa$-$\mu$ fading model in~\cite{6594884}\footnote{While the pioneering work presented in~\cite{6594884} refers to this model as $\kappa$-$\mu$ shadowed, to maintain consistency with the terminology adopted here we refer to it as an example of the single shadowed $\kappa$-$\mu$ Type I model.} and~\cite{cotton2015human}.
Accordingly, the PDF of $R$ is obtained as
\begin{equation}
f_{R}\left(r\right)=\frac{2m_{d}^{m_{d}}\left(1+\kappa\right)^{\mu}\mu^{\mu}}{\Gamma\left(\mu\right)\left(m_{d}+\kappa\mu\right)^{m_{d}}}\frac{r^{2\mu-1}}{\hat{r}^{2\mu}}\mathrm{e^{-\frac{r^{2}\left(1+\kappa\right)\mu}{\hat{r}^{2}}}} {}_{1}F_{1}\left(m_{d};\mu;\frac{\mu^{2}\kappa\left(1+\kappa\right)r^{2}}{\hat{r}^{2}\left(m_{d}+\kappa\mu\right)}\right)
\label{eq_section2_02}
\end{equation}
where, $\kappa > 0$ is the ratio of the total power of the dominant component  (${d}^2$) to that of the scattered waves ($2\mu{{\sigma}^2}$), $\mu > 0$ is related to the number of clusters, $\hat{r} =\sqrt{\mathbb{E}[R^{2}]}$ represents the rms power of $R$, the mean signal power is given by $\ensuremath{\mathbb{E}[R^{2}]} = 2\mu\sigma^{2}+d^{2}$, $\Gamma(\cdot)$ represents the gamma function and ${}_{1}F_{1}\left(\cdot;\cdot;\cdot\right)$ denotes the confluent hypergeometric function~\cite[eq. 9.210.1]{TofI}. 

 \setlength\parindent{35pt}{\textit{Example 2)}}
Our second example of the Type~I model assumes that the dominant component of a $\kappa$-$\mu$ signal undergoes variations influenced by an inverse Nakagami-$m$ RV. Thus, in \eqref{eq_section2_01} $\xi$ represents an inverse Nakagami-$m$ RV with shape parameter $m_d$ and $\mathbb{E}\left[{\xi}^2 \right] = 1$. The PDF of $R$ for this example case can be obtained via Theorem~\ref{thm Typethm1} below.
\begin{thm} For $\kappa$, $\mu$, $\hat{r}$ $\in\mathbb{R}^{+}$ and $m_d > 1$, the PDF of the single shadowed $\kappa$-$\mu$ Type I fading model for the example case when $\xi$ follows an inverse Nakagami-$m$ RV is expressed as
\begin{equation}
f_{R}\left(r\right)=\sum_{i=0}^{\infty}\frac{4\left[(m_{d}-1)\kappa)\right]^{\frac{m_{d}+i}{2}}r^{2i+2\mu-1}\mu^{\frac{1}{2}\left(3i+m_{d}\right)+\mu}}{\hat{r}^{2i+2\mu}~i!~\Gamma\left(m_{d}\right)\Gamma\left(i+\mu\right) \left(1+\kappa\right)^{-i-\mu}} \mathrm{e}^{\frac{-r^{2}\left(1+\kappa\right)\mu}{\hat{r}^{2}}}\mathrm{K}_{-i+m_{d}}\left(2\sqrt{(m_{d}\!-\!1)\mu\kappa}\right)
\label{eq_section2_new1}
\end{equation}
where ${\mathrm{K}_{\nu}(\cdot)}$ denotes the modified Bessel function of the second kind~\cite[eq. 9.6]{Abramowitz1972}.
\label{thm Typethm1}
\end{thm}
\begin{IEEEproof}
 See Appendix \ref{app:A0}.
\end{IEEEproof}

%%%%%%%%%%%%%%%%%%%%%%%%%%%%%%%%%%%%%%%%%%%%%%%%%%%%%%%%%%%
\subsection{Single Shadowed $\kappa$-$\mu$ Type~II Model}
%%%%%%%%%%%%%%%%%%%%%%%%%%%%%%%%%%%%%%%%%%%%%%%%%%%%%%%%%%%

The single shadowed $\kappa$-$\mu$ Type~II fading model assumes that the scattered waves in each cluster can randomly fluctuate because of shadowing. Its signal envelope, $R$, can be formulated in terms of the in-phase and quadrature phase components as

\begin{equation}
R^{2}=\sum_{i=1}^{\mu}\left(\xi X_{i}+ p_{i}\right)^{2}+\left(\xi Y_{i}+ q_{i}\right)^{2}
\label{eq_section2_03}
\end{equation}
where $\xi$, $X_i$, $Y_i$,  $p_i$, $q_i$ and $\mu$ are as defined previously. We now consider two example cases for the single shadowed Type II model, the details of which are discussed next. 

\setlength\parindent{35pt}{\textit{Example 1)}} In our first example of the single shadowed $\kappa$-$\mu$ Type~II model, we assume that the scattered components of a $\kappa$-$\mu$ signal undergo variations induced by a Nakagami-$m$ RV. Thus, in \eqref{eq_section2_03} $\xi$ denotes a Nakagami-$m$ RV with shape parameter $m_{s}$ and $\mathbb{E}\left[{\xi}^2 \right] = 1$. The PDF of $R$ for this example case can be obtained via Theorem~\ref{thm Type2a} below.

\begin{thm} For $\kappa$, $\mu$, $m_s$, $\hat{r}$ $\in\mathbb{R}^{+}$, the PDF of the single shadowed $\kappa$-$\mu$ Type II fading model for the example case when $\xi$ follows a Nakagami-$m$ RV can be expressed as

\begin{equation}
\!\!f_{R}\!\left(r\right)\!=\!\!\sum_{i=0}^{\infty}\!\frac{4\left(m_{s}\mu\right)^{\frac{1}{2}\left(2i+m_{s}+\mu\right)}r^{2i+2\mu-1}\kappa^{i}\left(1\!+\!\kappa\right)^{i+\mu}}{i!\Gamma\!\left(m_{s}\right)\Gamma\!\left(i\!+\!\mu\right)\left(r^{2}\left(1\!+\!\kappa\right)\!+\!\hat{r}^{2}\kappa\right)^{\frac{1}{2}\left(2i-m_{s}+\mu\right)}} \frac{1}{\hat{r}^{m_{s}+\mu}}\mathrm{K}_{2i-m_{s}+\mu}\!\!\left(\!\frac{2\sqrt{m_{s}\mu\left(r^{2}\left(1\!+\!\kappa\right)\!+\!\hat{r}^{2}\kappa\right)}}{\hat{r}}\right)\!\!.
\label{eq_section2_04}
\end{equation}
\label{thm Type2a}
\end{thm}
\vspace{-1cm}
\begin{IEEEproof}
 See Appendix~B. 
%\ref{app:A}.
\end{IEEEproof}
It can be seen from~\eqref{eq_section2_04} that the derivation of a closed-form expression for this example was infeasible. This was due to the inherent mathematical complexity of the resulting integral in~\eqref{eq_app_a5}. However, this is not the case with the second example of the single shadowed $\kappa$-$\mu$ Type~II model considered in this paper, whose PDF is derived next.

\setlength\parindent{35pt}{\textit{Example 2)}} 
Our second example of the Type~II model assumes that the scattered components of a $\kappa$-$\mu$ signal undergo variations induced by an inverse Nakagami-$m$ RV. Thus, in~\eqref{eq_section2_03} $\xi$ denotes an inverse Nakagami-$m$ RV with shape parameter $m_{s}$ and $\mathbb{E}\left[{\xi}^2 \right] = 1$. The PDF of $R$ for this example case can be obtained via Theorem~\ref{thm Type3a} as follows.
\begin{thm} For $\kappa$, $\mu$, $\hat{r}$ $\in\mathbb{R}^{+}$ and $m_s > 1$, the PDF of the single shadowed $\kappa$-$\mu$ Type~II fading model for the example case when $\xi$ follows an inverse Nakagami-$m$ RV is expressed as

\begin{align}
&f_{R}\left(r\right)=\frac{2(m_{s}-1)^{m_{s}}\left(1+\kappa\right)^{\mu}\mu^{\mu}r^{2\mu-1}\hat{r}^{2m_{s}}}{\mathrm{B}\left(m_{s},\mu\right)\left[r^{2}\left(1+\kappa\right)\mu+\hat{r}^{2}\left(m_{s} - 1 +\kappa\mu\right)\right]^{m_{s}+\mu}}\nonumber \\ &\!\!\times \!{}_{2}F_{1}\!\!\left(\frac{m_{s}\!+\!\mu}{2},\frac{1\!+\!m_{s}\!+\!\mu}{2};\mu;\frac{4\mu^{2}\kappa\left(1\!+\!\kappa\right)r^{2}\hat{r}^{2}}{\left[r^{2}\left(1+\kappa\right)\mu+\hat{r}^{2}\left(m_{s} - 1 +\kappa\mu\right)\right]^{2}}\!\right)
\label{eq_section2_06a}
\end{align}
where $\mathrm{B}(\cdot,\cdot)$ represents the Beta function~\cite[eq. 8.384]{TofI} and $_{2}F_{1}\left(\cdot,\cdot;\cdot;\cdot\right)$ denotes the Gauss hypergeometric function~\cite[eq. 9.100]{TofI}.
\label{thm Type3a}
\end{thm}
\begin{IEEEproof}
 See Appendix~B. 
%\ref{app:A}.
\end{IEEEproof}

%%%%%%%%%%%%%%%%%%%%%%%%%%%%%%%%%%%%%%%%%%%%%%%%%%%%%%%%%%%
\subsection{Single Shadowed $\kappa$-$\mu$ Type~III Model}
%%%%%%%%%%%%%%%%%%%%%%%%%%%%%%%%%%%%%%%%%%%%%%%%%%%%%%%%%%%
The single shadowed $\kappa$-$\mu$ Type~III fading model assumes that the rms power of a $\kappa$-$\mu$ signal can randomly fluctuate because of shadowing. Its signal envelope, $R$, can be formulated in terms of the in-phase and quadrature phase components as 
 
\begin{equation}
R^{2}= {\xi^{2}} \sum_{i=1}^{\mu}\left(X_{i}+ p_{i}\right)^{2}+\left(Y_{i}+ q_{i}\right)^{2}
\label{eq_section2_06}
\end{equation}
where, $\xi$, $X_i$, $Y_i$,  $p_i$, $q_i$ and $\mu$ are as defined previously. We now consider two example cases for the single shadowed Type~III model, the details of which are discussed next.

\setlength\parindent{35pt}{\textit{Example 1)}}
In our first example of the single shadowed $\kappa$-$\mu$ Type~III model, we assume that the multipath waves (both the dominant component and scattered waves) are subject to variations induced by a Nakagami-$m$ RV. Thus, in~\eqref{eq_section2_06} $\xi$ represents a Nakagami-$m$ RV with shape parameter $m_{t}$ and $\mathbb{E}\left[{\xi}^2 \right] = 1$. The PDF of $R$ for this example case is given by Theorem~\ref{thm Type5pdf}.
\begin{thm} For $\kappa$, $\mu$, $m_t$, $\hat{r}$ $\in\mathbb{R}^{+}$, the PDF of the single shadowed $\kappa$-$\mu$ Type~III fading model for the example case when $\xi$ follows a Nakagami-$m$ RV can be expressed as

\begin{equation}
f_{R}\left(r\right)=\sum_{i=0}^{\infty}\frac{4\left(m_{t}\mu\left(1+\kappa\right)\right)^{\frac{1}{2}\left(m_{t}+\mu+i\right)}\left(\kappa\mu\right)^{i}r^{m_{t}+\mu+i-1}}{\mathrm{e^{\kappa\mu}}i!\Gamma\left(m_{t}\right)\Gamma\left(i+\mu\right)\hat{r}^{m_{t}+\mu+i}} \mathrm{K}_{-m_{t}+\mu+i}\left(\frac{2r\sqrt{m_{t}\mu\left(1+\kappa\right)}}{\hat{r}}\right).
\label{eq_section2_15}
\end{equation}
where, $\kappa$, $\mu$, $\hat{r}$ are as defined previously.
\label{thm Type5pdf}
\end{thm}
\begin{IEEEproof}
 See Appendix~C. 
%\ref{app:C}.
\end{IEEEproof}
\noindent Note that it is also possible to derive this PDF as a special case of the statistics of the product of $\kappa$-$\mu$ and Nakagami-$m$ RVs as shown in~\cite{carlos2017} and~\cite{newTWireless}. 

{\setlength\parindent{35pt}{\textit{Example 2)}} Our second example of the Type~III model assumes that the multipath waves are subject to variations induced by an inverse Nakagami-$m$ RV. Thus, in~\eqref{eq_section2_06} $\xi$ represents an inverse Nakagami-$m$ RV with shape parameter $m_{t}$ and $\mathbb{E}\left[{\xi}^2 \right] = 1$.} This example of the single shadowed $\kappa$-$\mu$ Type~III model was introduced in~\cite{newTCOM} as the $\kappa$-$\mu$/inverse gamma fading model in which the mean power of the multipath waves were subject to fluctuations induced by an inverse gamma RV. 
Furthermore, the PDF of this model can be obtained as a special case of the statistics of the ratio of $\kappa$-$\mu$ and inverse Nakagami-$m$ RVs as shown in~\cite{carlosratio}. Since the inverse Nakagami-$m$ RV used for this analysis is assumed to have $\mathbb{E}\left[{\xi}^2 \right] = 1$, the PDF of $R$ for the single shadowed $\kappa$-$\mu$ Type~III fading model for this example case can be obtained by substituting $\hat{r}^{2} = \frac{(m_{t}-1)\hat{r}^{2}}{m_{t}}$ in~\cite{newTCOM}, which yields

\begin{equation}
f_{R}\left(r\right) =\frac{\mathrm{2\left(1+\kappa\right)^{\mu}\mu^{\mu}e^{-\kappa\mu}}\left((m_{t}-1)\hat{r}^{2}\right)^{m_{t}}r^{2\mu-1}}{{\rm{B}}\!\left(m_{t},\mu\!\right)\left(\hat{r}^{2}(m_{t}-1)+r^{2}\left(1+\kappa\right)\mu\right)^{m_{t}+\mu}} {}_{1}F_{1}\left(m_{t}+\mu;\mu;\frac{\mu^{2}\kappa\left(1+\kappa\right)r^{2}}{\hat{r}^{2}(m_{t}-1)+r^{2}\left(1+\kappa\right)\mu}\right)
\label{eq_section2_08} 
\end{equation}
where $m_{t} > 1$.

 %The PDF of $R$ for this example case is given by Theorem~\ref{thm Type5pdf}.} 
%
%{\color{red}Similar to the single shadowed $\kappa$-$\mu$ Type~I~and~II models, two example cases are discussed for the single shadowed $\kappa$-$\mu$ Type III model.} In particular, the {first example of the \color{red}Type~III} model considers that the multipath waves (both the dominant component and scattered waves) are subject to variations induced by a Nakagami-$m$ RV, whilst the {\color{red}second example}  assumes that this variation is brought about by an inverse Nakagami-$m$ RV. Accordingly, their signal envelope $R$ can be formulated in terms of the {\color{red}in-phase and quadrature phase} components as 
%
%
%{\color{red}\subsubsection{Example A}
%In the first example of the single shadowed Type~III model, $\xi$ in~\eqref{eq_section2_06} represents a Nakagami-$m$ RV  with shape parameter $m_{t}$ and $\mathbb{E}\left[{\xi}^2 \right] = 1$.  
%
%{\color{red} \subsubsection{Example B}
%In the second example of the single shadowed Type~III model considered here, we assume that $\xi$ in~\eqref{eq_section2_06} represents an inverse Nakagami-$m$ RV with shape parameter $m_{t}$ and $\mathbb{E}\left[{\xi}^2 \right] = 1$.} 
%Note that {\color{red}this example of the} single shadowed $\kappa$-$\mu$ {\color{red}Type~III} model was introduced in~\cite{newTCOM} as the $\kappa$-$\mu$/inverse gamma fading model in which the mean power of the multipath waves were subject to fluctuations induced by an inverse gamma RV. 

%%%%%%%%%%%%%%%%%%%%%%%%%%%%%%%%%%%%%%%%%%%%%%%%%%%%%%%%%%%
\section{Double Shadowed $\kappa$-$\mu$ Models}
%%%%%%%%%%%%%%%%%%%%%%%%%%%%%%%%%%%%%%%%%%%%%%%%%%%%%%%%%%%
In this section, we discuss three different ways in which the $\kappa$-$\mu$ fading envelope can be impacted by more than one shadowing process. To this end, we propose the double shadowed $\kappa$-$\mu$ Type~I to Type~III fading models with their physical interpretation provided in Table~II.

%%%%%%%%%%%%%%%%%%%%%%%%%%%%%%%%%%%%%%%%%%%%%%%%%%%%%%%%%%%
\subsection{Double Shadowed $\kappa$-$\mu$ Type~I Model}
%%%%%%%%%%%%%%%%%%%%%%%%%%%%%%%%%%%%%%%%%%%%%%%%%%%%%%%%%%%
The double shadowed $\kappa$-$\mu$ Type~I model characterizes the propagation scenario in which the envelope experiences shadowing of the dominant component, which is preceded (or succeeded) by a secondary round of multiplicative shadowing. Physically, this situation may arise when the signal power delivered through the optical path between the transmitter and receiver is shadowed by objects moving within its locality, whilst further shadowing of the received power (combined multipath and dominant paths) may also occur due to obstacles moving in the vicinity of the transmitter or receiver. Following from this, its signal envelope, $R$, can be expressed in terms of the in-phase and quadrature phase components as 

\begin{equation}
R^{2}=A^2\sum_{i=1}^{\mu}\left(X_{i}+\xi p_{i}\right)^{2}+\left(Y_{i}+\xi q_{i}\right)^{2}
\label{eq_section3_01}
\end{equation}
where $\xi$, $\mu$, $X_i$, $Y_i$, $p_i$ and $q_i$ are as defined previously and $A$ represents a RV which introduces an additional degree of shadowing. As before, we now provide two example cases of the double shadowed $\kappa$-$\mu$ Type~I model.

\setlength\parindent{35pt}{\textit{Example 1)}}
In our first example of the double shadowed $\kappa$-$\mu$ Type~I model, we assume that the shadowing of the dominant component is shaped by a Nakagami-$m$ RV, whilst the second round of multiplicative shadowing is induced by an inverse Nakagami-$m$ RV. Thus, in~\eqref{eq_section3_01} $\xi$  represents a Nakagami-$m$ RV (with shape parameter $m_d$ and $\mathbb{E}\left[{\xi}^2 \right] = 1$) whilst $A$ denotes an inverse Nakagami-$m$ RV (with shape parameter $m_t$ and $\mathbb{E}\left[{A}^2 \right] = 1$).
Accordingly, the PDF of the double shadowed $\kappa$-$\mu$ Type~I fading model for this example case{\footnote{Note that this model was also introduced in~\cite{doubleshadowedconference} (as early results of this work) as a new fading model which is capable of characterizing both the shadowing of the dominant component and composite shadowing which may exist in wireless channels.}} can be obtained via Theorem~\ref{thm Type1apdf}.
\begin{thm} For $\kappa$, $\mu$, $m_d$, $\hat{r}$ $\in\mathbb{R}^{+}$ and $m_t > 1$, the PDF of the double shadowed $\kappa$-$\mu$ Type~I fading model for the example case when $\xi$ represents a Nakagami-$m$ RV and $A$ represents an inverse Nakagami-$m$ RV is expressed as

\begin{equation}
\!\!\!f_{R}\!\left(r\right)\!=\!\frac{2(m_{t}-1)^{m_{t}}m_{d}^{m_{d}}\mathcal{K}^{\mu}r^{2\mu-1}\hat{r}^{2m_{t}} {}_{2}F_{1}\left(m_{d},m_{t}\!+\!\mu;\mu;\frac{\mathcal{K}\mu\kappa r^{2}}{\left(m_{d}+\mu\kappa\right)\left(\mathcal{K}r^{2}+(m_{t}-1)\hat{r}^{2}\right)}\right)}{\left(m_{d}+\mu\kappa\right)^{m_{d}} {\rm{B}}\left(m_{t},\mu\right)\left(\mathcal{K}r^{2}+(m_{t}-1)\hat{r}^{2}\right)^{m_{t}+\mu}}
\label{eq_section3_02}
\end{equation}
where $\mathcal{K}=\mu\left(1+\kappa\right)$.
\label{thm Type1apdf}
\end{thm}
\begin{IEEEproof}
 See Appendix~D. 
%\ref{app:D}.
\end{IEEEproof}

\setlength\parindent{35pt}{\textit{Example 2)}}
Our second example of the double shadowed Type~I model assumes that the shadowing of the dominant component is brought about by an inverse Nakagami-$m$ RV, whilst the second round of multiplicative shadowing is influenced by a Nakagami-$m$ RV. Thus, in~\eqref{eq_section3_01} $\xi$  represents an inverse Nakagami-$m$ RV (with shape parameter $m_d$ and $\mathbb{E}\left[{\xi}^2 \right] = 1$) whilst $A$ denotes an Nakagami-$m$ RV (with shape parameter $m_t$ and $\mathbb{E}\left[{A}^2 \right] = 1$). The PDF of the double shadowed $\kappa$-$\mu$ Type~I fading model for this example case can be obtained via Theorem~\ref{thm TypeIIIpdf}.

%{\color{red} \subsubsection{Example B}  
%In the second example of the double shadowed $\kappa$-$\mu$ Type~I model we consider that the shadowing of the dominant component is brought about by an inverse Nakagami-$m$ RV whilst the secondary round of multiplicative shadowing is shaped by a Nakagami-$m$ RV. Thus, $\xi$ (see~\eqref{eq_section3_01}) represents an inverse Nakagami-$m$ RV with shape parameter $m_d$ and $\mathbb{E}\left[{\xi}^2 \right] = 1$ whereas $A$ denotes a Nakagami-$m$ RV with shape parameter $m_t$ and $\mathbb{E}\left[{A}^2 \right] = 1$. The PDF of the double shadowed $\kappa$-$\mu$ {\color{red}Type~I} fading model for this example case can be obtained via Theorem~\ref{thm TypeIIIpdf}.}

	\begin{thm} For $\kappa$, $\mu$, $m_t$, $m_d$, $\hat{r}$ $\in\mathbb{R}^{+}$, the PDF of the double shadowed $\kappa$-$\mu$ Type~I fading model when $\xi$ represents an inverse Nakagami-$m$ RV and $A$ represents a Nakagami-$m$ RV can be expressed as
			\begin{align}
			f_{R}\left(r\right) &= \frac{8 \left(m_t \mathcal{K} \right)^{\frac{\mu +m_t}{2} }  r^{\mu +m_t-1}}{\Gamma \left(m_d\right) \Gamma \left(m_t\right) \left(\kappa  \mu  (m_d -1)\right)^{-\frac{m_d}{2}}	\hat{r}^{\mu +m_t}} \sum _{i=0}^{\infty } \frac{1}{i! \Gamma (i+\mu )}
		\left(\frac{r \mu  \sqrt{\kappa  (m_d -1) m_t (1+\kappa )}}{\hat{r}}\right)^i \nonumber \\
			&\times {\rm{K}}_{m_d-i}\left(2 \sqrt{(m_d -1) \mu  \kappa }\right) {\rm{K}}_{m_t-\mu -i}\left(\frac{2 r \sqrt{\mathcal{K} m_t}}{\hat{r}}\right). 
			\label{eq_section3_10}
			\end{align}
			\label{thm TypeIIIpdf}
			\vspace{-1cm}
		\end{thm}
		\begin{IEEEproof}
			See Appendix~E. 
			%\ref{app:H}.
		\end{IEEEproof}

%%%%%%%%%%%%%%%%%%%%%%%%%%%%%%%%%%%%%%%%%%%%%%%%%%%%%%%%%%%
\subsection{Double Shadowed $\kappa$-$\mu$ Type~II Model}
%%%%%%%%%%%%%%%%%%%%%%%%%%%%%%%%%%%%%%%%%%%%%%%%%%%%%%%%%%%
The double shadowed $\kappa$-$\mu$ Type~II model considers a $\kappa$-$\mu$ faded signal in which the dominant component and scattered waves experience two different shadowing processes. Its signal envelope, $R$, is given by

\begin{equation}
R^{2}=\sum_{i=1}^{\mu}\left(AX_{i}+ Bp_{i}\right)^{2}+\left(AY_{i}+ Bq_{i}\right)^{2}
\label{eq_section3_07}
\end{equation}
where $\mu$, $X_i$, $Y_i$, $p_i$ and $q_i$ are as defined previously; $A$ and $B$ represent RVs that are responsible for introducing two different shadowing processes. We now consider two example cases for the double shadowed $\kappa$-$\mu$ Type~II model, the details of which are discussed next.

\setlength\parindent{35pt}{\textit{Example 1)}}
In our first example of the double shadowed Type~II model, we assume that the dominant component of a $\kappa$-$\mu$ signal undergoes variations influenced by a Nakagami-$m$ RV, whilst the scattered waves of a $\kappa$-$\mu$ signal are subject to variations induced by an inverse Nakagami-$m$ RV. Thus, in~\eqref{eq_section3_07} $A$ denotes an inverse Nakagami-$m$ RV with shape parameter $m_s$, and $B$ represents a Nakagami-$m$ RV with shape parameter $m_d$. Here, $\mathbb{E}\left[{A}^2 \right]$ and $\mathbb{E}\left[{B}^2 \right]$ are set equal to 1.  
%{\color{red}\subsubsection{Example A}
%In the first example of the double shadowed Type~II model, $A$ in~\eqref{eq_section3_07} denotes an inverse Nakagami-$m$ RV with shape parameter $m_s$, and $B$ (see~\eqref{eq_section3_07}) represents a Nakagami-$m$ RV with shape parameter $m_d$. Here, $\mathbb{E}\left[{A}^2 \right]$ and $\mathbb{E}\left[{B}^2 \right]$ are set equal to 1.} 
An analytical expression for the PDF of the double shadowed Type~II fading model for this example case can be obtained via Theorem~\ref{thm Type2pdf} below.
 \begin{thm} For $\kappa$, $\mu$, $m_d$, $\hat{r}$ $\in\mathbb{R}^{+}$, and $m_s > 1$ the PDF of the double shadowed $\kappa$-$\mu$ Type~II fading model when $A$ denotes an inverse Nakagami-$m$ RV and $B$ denotes a Nakagami-$m$ RV can be expressed as

\begin{align}
&f_{R}\left(r\right) =\frac{2 \left(m_s-1\right){}^{m_s} m_d^{m_s+\mu } r^{2 \mu -1} \Gamma \left(\mu +m_s\right) (1+\kappa
	)^{\mu }}{\kappa ^{m_s+\mu } \Gamma \left(m_s\right) \Gamma \left(m_d\right) \mu ^{m_s} \hat{r}^{2
		\mu }}
\sum _{i=0}^{\infty } \frac{2^{2 i} \left(\frac{\mu +m_s}{2} \right)_i \left(\theta	_1\right)_i}{i! \Gamma (\mu +i)} \Gamma \left(i+m_d\right) \nonumber \\
&\times \left(\frac{m_d (1+\kappa ) r^2}{\hat{r}^2 \kappa }\right)^i  {\rm{U}}\left(2 i+\mu +m_s,1+i+\mu -m_d+m_s,\theta _2\right)
\label{eq_section3_08}
\end{align}
where $\theta_1 = \frac{1}{2}\left(1+m_{s}+\mu\right) $, $\theta_2 = \frac{m_{d}\left((m_{s}-1)\hat{r}^{2}+r^{2}\left(1+\kappa\right)\mu\right)}{\hat{r}^{2}\kappa\mu}$, $(a)_i$ is the Pochhammer's symbol \cite[eq. 6.1.22]{Abramowitz1972} and $\mathrm{U}(\cdot,\cdot, \cdot)$ is the confluent Tricomi hypergeometric function~\cite[eq. 13.1.3]{Abramowitz1972}
\label{thm Type2pdf}
\end{thm}
\begin{IEEEproof}
 See Appendix~F. 
%\ref{app:F}.
\end{IEEEproof}

\setlength\parindent{35pt}{\textit{Example 2)}}
Our second example of the double shadowed Type~II model assumes that the dominant component of a $\kappa$-$\mu$ signal undergoes variations influenced by an inverse Nakagami-$m$ RV whilst the scattered waves of a $\kappa$-$\mu$ signal are subject to variations induced by a Nakagami-$m$ RV. Thus, in~\eqref{eq_section3_07} $A$ denotes a Nakagami-$m$ RV (with shape parameter $m_s$ and $\mathbb{E}\left[{A}^2 \right] = 1$), and $B$ represents an inverse Nakagami-$m$ RV (with shape parameter $m_d$ and $\mathbb{E}\left[{B}^2 \right] = 1$).
%{\color{red}\subsubsection{Example B}
%In the second example of the double shadowed Type~II model, $A$ in~\eqref{eq_section3_07} denotes a Nakagami-$m$ RV (with shape parameter $m_s$ and $\mathbb{E}\left[{A}^2 \right] = 1$), and $B$ (see~\eqref{eq_section3_07}) represents an inverse Nakagami-$m$ RV (with shape parameter $m_d$ and $\mathbb{E}\left[{B}^2 \right] = 1$).} 
The PDF of the double shadowed Type~II model for this example case can be obtained via Theorem~\eqref{thm TypeVIpdf} as follows.		

	\begin{thm} For $\kappa$, $\mu$, $m_s$, $\hat{r}$ $\in\mathbb{R}^{+}$, and $m_d > 1$ the PDF of the double shadowed $\kappa$-$\mu$ Type~II fading model when $A$ denotes a Nakagami-$m$ RV and $B$ represents an inverse Nakagami-$m$ RV can be expressed as
		\begin{equation}
\!\!\!f_{R}\left(r\right)=\frac{4 \pi   \left(\mathcal{K} m_s\right)^{\frac{1}{2} \left(\mu +m_s\right)} r^{\mu +m_s-1}}{\sin \left(\pi  m_d\right)\Gamma \left(m_d\right) \Gamma \left(m_s\right) \hat{r}^{\mu +m_s}} \!\!\sum _{i=0}^{\infty } \frac{1}{i!}
		\Bigg(\!\frac{\mathcal{M}}{\left(i\!-\!m_d\right)!}\! \left(\!\frac{\mathcal{P}}{r \sqrt{\mathcal{K}}}\right)^{\!\!i}
		-\frac{\!\mathcal{N}}{\left(i\!+\!m_d\right)!} \!\!\left(\!\frac{\mathcal{P}}{r \sqrt{\mathcal{K}}}\right)^{\!\!i+m_d}\Bigg)
		\label{eq_section3_14}
		\end{equation}
where $\mathcal{K}$ is as defined previously, $\mathcal{P} = \kappa  \mu  \hat{r} (m_d-1) \sqrt{m_s}$,

\begin{align}
&\!\!\!\!\!\mathcal{M} = {\rm{K}}_{i+\mu -m_s}\left(\frac{2 r \sqrt{\mathcal{K} m_s}}{\hat{r}}\right) \, _3\tilde{F}_1\left(-\frac{i}{2},\frac{1-i}{2},m_d-i;\mu ;\frac{4 r^2 \mathcal{K}}{\kappa  \mu  \hat{r}^2 (m_d-1)}\right),\\
&\!\!\!\!\!\mathcal{N}={\rm{K}}_{i+\mu +m_d-m_s}\left(\frac{2 r \sqrt{\mathcal{K} m_s}}{\hat{r}}\right) \, _3\tilde{F}_1\left(-i,\frac{1}{2} \left(-i\!-\!m_d\right),\frac{1}{2} \left(1\!-\!i\!-\!m_d\right);\mu ;\frac{4 r^2
	\mathcal{K}}{\kappa  \mu  \hat{r}^2 (m_d\!-\!1)}\right)
	\end{align}
	and $_3\tilde{F}_1\left(\cdot, \cdot, \cdot; \cdot;\cdot\right)$ is the generalized hypergeometric function~\cite{prudinkov_v3}.
			\label{thm TypeVIpdf}
	\end{thm}
	\begin{IEEEproof}
		See Appendix~G. 
		%\ref{app:J}
	\end{IEEEproof}

%in that the dominant and scattered components are fluctuated by different shadowing processes.
For conciseness, it is worth mentioning here that two further examples of the double shadowed model can readily be obtained from~\eqref{eq_section3_07}, which coincidentally lead to PDFs equivalent in form to those given in~\eqref{eq_section3_02} and~\eqref{eq_section3_10}. These can be found by letting $B = A\xi$, where $A$ and $\xi$ represent either a Nakagami-$m$ and an inverse Nakagami-$m$ RV or vice versa. It is worth highlighting that as shown in~\cite{7886273}, $B^2$ follows a Fisher-Snedecor $\mathcal{F}$ distribution~\cite{walck1996hand}. Now, substituting for $B$ in~\eqref{eq_section3_07} we evidently arrive at~\eqref{eq_section3_01}. Then letting $A$ denote an inverse Nakagami-$m$ RV and $\xi$ represent a Nakagami-$m$ RV and following the same statistical procedure highlighted in Section~III.A, the PDF in~\eqref{eq_section3_02} is deduced. Similarly, if we let $A$ denote a Nakagami-$m$ RV and $\xi$ represent an inverse Nakagami-$m$ RV, we arrive at~\eqref{eq_section3_10}.

	%%%%%%%%%%%%%%%%%%%%%%%%%%%%%%%%%%%%%%%%%%%%%%%%%%%%%%%%%%%
	\subsection{Double Shadowed $\kappa$-$\mu$ Type III Model}
	%%%%%%%%%%%%%%%%%%%%%%%%%%%%%%%%%%%%%%%%%%%%%%%%%%%%%%%%%%%
The double shadowed $\kappa$-$\mu$ Type~III fading model considers a $\kappa$-$\mu$ faded signal in which the scattered waves in each cluster are subject to fluctuations caused by shadowing. As well as this, it assumes that the rms power of the dominant component and scattered waves may also be subject to random variations induced by shadowing. Its signal envelope, $R$ is expressed as

		\begin{equation}\label{eq_section3_11}
		R^2 = A^2 \sum\limits_{i=1}^\mu (\xi X_i+p_i)^2 + (\xi Y_i+q_i)^2
		\end{equation}
		where $\xi$, $A$, $\mu$, $X_i$, $Y_i$, $p_i$, and $q_i$ are defined previously. 	
As before, we now provide two example cases for the double shadowed Type~III model.

\setlength\parindent{35pt}{\textit{Example 1)}}
 In our first example of the double shadowed $\kappa$-$\mu$ Type~III model, we assume that the shadowing of the scattered components is influenced by an inverse Nakagami-$m$ RV whilst the secondary round of multiplicative shadowing is induced by a Nakagami-$m$ RV. Thus, in~\eqref{eq_section3_11} $A$ denotes a Nakagami-$m$ RV (with shape parameter $m_t$ and $\mathbb{E}\left[{A}^2 \right] = 1$) whilst $\xi$ represents an inverse Nakagami-$m$ RV
(with shape parameter $m_s$ and $\mathbb{E}\left[{\xi}^2 \right] = 1$). The PDF of the double shadowed Type~III model for this example can be obtained via Theorem~\ref{thm TypeIVpdf} below.
\begin{thm} For $\kappa$, $\mu$, $m_t$, $\hat{r}$ $\in\mathbb{R}^{+}$, and $m_s > 1$ the PDF of the double shadowed $\kappa$-$\mu$ Type~III fading model where $A$ denotes a Nakagami-$m$ RV and $\xi$ denotes an inverse Nakagami-$m$ RV is given by

			\begin{align}
			f_{R}\left(r\right) &=\! \frac{2 \left(m_s-1\right){}^{m_s} \left(m_t \mathcal{K}\right){}^{\mu } r^{2 \mu -1}}{\Gamma \left(m_t\right) B\left(m_s,\mu \right) \hat{r}^{2 \mu } \left(m_s-1+\kappa  \mu \right)^{\mu +m_s}}
			\sum _{i=0}^{\infty } \frac{\left(\frac{1}{2} \left(m_s+\mu \right)\right)_i \left(\theta _1\right)_i \Gamma \left(i+m_s+m_t\right)}{i! (\mu )_i \hat{r}^{2 i}}\nonumber \\
			&\times \frac{\left(4 r^2 \kappa  \mu  \mathcal{K} m_t\right)^i}{\left(m_s-1+\kappa  \mu \right)^{2 i}} {\rm{U}}\left(2 i+\mu +m_s,1+i+\mu -m_t,\frac{r^2 \mathcal{K} m_t}{\hat{r}^2 \left(m_s-1+\kappa  \mu \right)}\right)
			 \label{eq_section3_12}
			\end{align}
			\label{thm TypeIVpdf}
			in which $\mathcal{K}$ and $\theta_1$ are defined previously.
			\vspace{-0.2cm}
		\end{thm}
		\begin{IEEEproof}
			See Appendix~H. 
			%\ref{app:I}.
		\end{IEEEproof}

\setlength\parindent{35pt}{\textit{Example 2)}}
Our second example of the double shadowed $\kappa$-$\mu$ Type~III model assumes that the shadowing of the scattered components is influenced by a Nakagami-$m$ RV whilst the secondary round of multiplicative shadowing is induced by an inverse Nakagami-$m$ RV. Thus, in~\eqref{eq_section3_11} $A$ denotes an inverse Nakagami-$m$ RV (with shape parameter $m_t$ and $\mathbb{E}\left[{A}^2 \right] = 1$) whilst $\xi$ represents a Nakagami-$m$ RV
(with shape parameter $m_s$ and $\mathbb{E}\left[{\xi}^2 \right] = 1$). The PDF of the double shadowed Type~III model for this example can be obtained via Theorem~\ref{thm TypeVpdf} as follows.
\begin{thm} For $\kappa$, $\mu$, $m_s$, $\hat{r}$ $\in\mathbb{R}^{+}$, and $m_t > 1$ the PDF of the double shadowed $\kappa$-$\mu$ Type~III fading model when $A$ denotes an inverse Nakagami-$m$ RV and $\xi$ denotes a Nakagami-$m$ RV is given by

	\begin{align}
	&f_{R}\left(r\right) = \frac{2 \pi  \left(\mathcal{K} m_s/(m_t-1)\right)^{m_s} r^{2 m_s-1}\hat{r}^{-2 m_s}}{\Gamma \left(m_t\right) \Gamma \left(m_s\right) \sin \left(\pi  \left(m_t+m_s\right)\right) } \sum _{i=0}^{\infty } \frac{\left(\kappa  \mu  m_s\right)^i}{i!}
 \Bigg(\mathcal{G} \left(\frac{\kappa  \mu  \hat{r}^2 (m_t-1)}{r^2 \mathcal{K}}\right)^{-i}
\nonumber \\ &-\mathcal{H} \left(\frac{\kappa  \mu  \hat{r}^2 (m_t-1)}{r^2 \mathcal{K}}\right)^{m_t+m_s} + \mathcal{J} \left(\frac{r^2
		\mathcal{K} m_s}{\hat{r}^2 (m_t-1)}\right)^{i+\mu -m_s} \left(\kappa  \mu  m_s\right)^{-i}\Bigg)
	\label{eq_section3_13}
	\end{align}
in which $\mathcal{K}$ is defined previously, and

\begin{align}
&	\!\!\!\!\!\!\mathcal{G}\!=\! \frac{(-1)^i \pi  \csc \left(\pi  \left(\mu\! -\!m_s\right)\right)}{\Gamma \left(1\!+\!i\!-\!\mu\! +\!m_s\right)} \, _2\tilde{F}_2\!\left(\!1\!-\!i\!-\!m_s,-\!i\!+\!\mu\! -\!m_s;\mu ,1\!-\!i\!-\!m_t\!-\!m_s;-\!\frac{\kappa  \mu  \hat{r}^2 (m_t\!-\!1)}{r^2 \mathcal{K}}\!\right),\\
&	\!\!\!\!\!\!\!\!\mathcal{H} \!=\! \frac{\Gamma\! \left(\mu\! +\!m_t\right) \Gamma\! \left(i\!+\!m_s\right)}{\Gamma \left(-m_t\right)} \, \! _2\tilde{F}_2\!\left(\!1\!+\!m_t,\mu\! +\!m_t;1\!+\!i\!+\!m_t\!+\!m_s,i\!+\!\mu\! +\!m_t\!+\!m_s;-\!\frac{\kappa  \mu  \hat{r}^2 (m_t\!-\!1)}{r^2 \mathcal{K}}\!\right),\\
&	\!\!\!\!\!\!\!\mathcal{J}  =  \frac{\Gamma \left(-\!i\!-\mu\! +\!m_s\right) \sin \left(\pi  \left(m_t\!+\!m_s\right)\right)}{\sin \left(\pi  \left(\mu\! +\!m_t\right)\right)} \, _2\tilde{F}_2\left(-i,1\!-\!i\!-\!\mu ;\mu ,1\!-\!i\!-\!\mu\! -\!m_t;\!-\frac{\kappa  \hat{r}^2 (m_t\!-\! 1)}{r^2 (1+\kappa )}\right),
	\end{align}
where $\, _2\tilde{F}_2(a,b;c,d,z) = \, _2F_2(a,b;c,d,z)/ (\Gamma(c)\Gamma(d))$ is a particular case of the generalized hypergeometric function \cite[eq. 7.2.3.1]{prudinkov_v3}.
\label{thm TypeVpdf}
\end{thm}
\begin{IEEEproof} See Appendix~I
%\ref{app:K}.
\end{IEEEproof}

\begin{table*} 
\small
		\renewcommand{\arraystretch}{1.5}
		  \captionsetup{justification=centering}
		\caption{Special Cases of the Double Shadowed $\kappa$-$\mu$ Type~I~(example~1), Type~I~(example~2) and Type~II~(example~1) Fading Models}
		\label{Table:1}
		\centering
		\begin{threeparttable}
		\begin{tabular}{|c|c|c|c|}
\hline 
\multirow{2}{*}{Fading models} & double shadowed $\kappa$-$\mu$  & double shadowed $\kappa$-$\mu$  & double shadowed $\kappa$-$\mu$ \tabularnewline
 & Type~I~(example~1) & Type~I~(example~2) & Type~II~(example~1)\tabularnewline
\hline 
\hline 
Single shadowed $\kappa$-$\mu$  & $\ensuremath{\underline{m_{t}}\rightarrow\infty},\ensuremath{\underline{m_{d}}=m_{d}},$ & \multirow{2}{*}{-} & $\ensuremath{\underline{m_{s}}\rightarrow\infty},\ensuremath{\underline{m_{d}}=m_{d}},$\tabularnewline
 Type~I~(example~1)~\cite{6594884} & $\ensuremath{\underline{\kappa}=\kappa},\ensuremath{\underline{\mu}=\mu}$ &  & $\ensuremath{\underline{\kappa}=\kappa},\ensuremath{\underline{\mu}=\mu}$\tabularnewline
\hline 
Single shadowed & \multirow{2}{*}{-} & $\ensuremath{\ensuremath{\underline{m_{t}}\rightarrow\infty},\ensuremath{\underline{m_{d}}=m_{d}},}$ & \multirow{2}{*}{-}\tabularnewline
$\kappa$-$\mu$ Type~I~(example~2) &  & $\ensuremath{\ensuremath{\underline{\kappa}=\kappa},\ensuremath{\underline{\mu}=\mu}}$ & \tabularnewline
\hline 
Single shadowed  & \multirow{2}{*}{-} & \multirow{2}{*}{-} & \multirow{2}{*}{-}\tabularnewline
$\kappa$-$\mu$ Type~II~(example~1) &  &  & \tabularnewline
\hline 
Single shadowed & \multirow{2}{*}{-} & \multirow{2}{*}{-} & $\ensuremath{\underline{m_{s}}\rightarrow m_{s}},\ensuremath{\underline{m_{d}}\rightarrow\infty},$\tabularnewline
$\kappa$-$\mu$ Type~II~(example~2) &  &  & $\ensuremath{\underline{\kappa}=\kappa},\ensuremath{\underline{\mu}=\mu}$\tabularnewline
\hline 
Single shadowed $\kappa$-$\mu$ & \multirow{2}{*}{-} & $\ensuremath{\ensuremath{\underline{m_{t}}\rightarrow m_{t}},\ensuremath{\underline{m_{d}}\rightarrow\infty},}$ & \multirow{2}{*}{-}\tabularnewline
Type~III~(example~1)~\cite{carlos2017} &  & $\ensuremath{\ensuremath{\underline{\kappa}=\kappa},\ensuremath{\underline{\mu}=\mu}}$ & \tabularnewline
\hline 
Single shadowed $\kappa$-$\mu$ & $\ensuremath{\underline{m_{t}}=m_{t}},\ensuremath{\underline{m_{d}}\rightarrow\infty},$ & \multirow{2}{*}{-} & \multirow{2}{*}{-}\tabularnewline
Type~III~(example~2) & $\ensuremath{\underline{\kappa}=\kappa},\ensuremath{\underline{\mu}=\mu}$ &  & \tabularnewline
\hline 
\multirow{3}{*}{$\eta-\mu$/inverse gamma~\cite{newTCOM}} & $\underline{m_{t}}\rightarrow\infty$, $\underline{m_{d}}\rightarrow\infty$ & \multirow{3}{*}{-} & \multirow{3}{*}{-}\tabularnewline
 & $\underline{\kappa}\rightarrow\frac{(1-\eta)}{2\eta},$$\underline{\mu}=2\mu$ &  & \tabularnewline
 & $\underline{\hat{r}}^{2}=\frac{m_{t}\hat{r}^{2}}{(m_{t}-1)}$ &  & \tabularnewline
\hline 
\multirow{2}{*}{$\kappa$-$\mu$} & $\ensuremath{\underline{m_{t}}\rightarrow\infty},\ensuremath{\underline{m_{d}}\rightarrow\infty},$ & $\ensuremath{\underline{m_{t}}\rightarrow\infty},\underline{m_{d}}\rightarrow\infty,$ & $\ensuremath{\underline{m_{s}}\rightarrow\infty},\ensuremath{\underline{m_{d}}\rightarrow\infty},$\tabularnewline
 & $\ensuremath{\underline{\kappa}=\kappa},\ensuremath{\underline{\mu}=\mu}$ & $\ensuremath{\underline{\kappa}=\kappa},\ensuremath{\underline{\mu}=\mu}$ & $\ensuremath{\underline{\kappa}=\kappa},\ensuremath{\underline{\mu}=\mu}$\tabularnewline
\hline 
\multirow{2}{*}{$\eta$-$\mu$} & $\ensuremath{\underline{m_{t}}\rightarrow\infty},\ensuremath{\underline{m_{d}}\rightarrow\mu},$ & \multirow{2}{*}{-} & $\ensuremath{\underline{m_{s}}\rightarrow\infty},\ensuremath{\underline{m_{d}}\rightarrow\mu},$\tabularnewline
 & $\ensuremath{\underline{\kappa}=\frac{(1-\mbox{\ensuremath{\eta}})}{2\eta}},\ensuremath{\underline{\mu}=2\mu}$ &  & $\ensuremath{\underline{\kappa}=\frac{(1-\mbox{\ensuremath{\eta}})}{2\eta}},\ensuremath{\underline{\mu}=2\mu}$\tabularnewline
\hline 
\multirow{2}{*}{Shadowed Rician~\cite{abdi2003new} } & $\ensuremath{\underline{m_{t}}\rightarrow\infty},\ensuremath{\underline{m_{d}}=m_{d}},$ & \multirow{2}{*}{-} & $\ensuremath{\underline{m_{s}}\rightarrow\infty},\ensuremath{\underline{m_{d}}=m_{d}},$\tabularnewline
 & $\ensuremath{\underline{\kappa}=k},\ensuremath{\underline{\mu}=1}$ &  & $\ensuremath{\underline{\kappa}=k},\ensuremath{\underline{\mu}=1}$\tabularnewline
\hline 
\multirow{2}{*}{Rician} & $\ensuremath{\underline{m_{t}}\rightarrow\infty},\ensuremath{\underline{m_{d}}\rightarrow\infty},$ & $\ensuremath{\underline{m_{t}}\rightarrow\infty},\ensuremath{\underline{m_{d}}\rightarrow\infty},$ & $\ensuremath{\underline{m_{s}}\rightarrow\infty},\ensuremath{\underline{m_{d}}\rightarrow\infty},$\tabularnewline
 & $\ensuremath{\underline{\kappa}=k},\ensuremath{\underline{\mu}=1}$ & $\ensuremath{\underline{\kappa}=k},\ensuremath{\underline{\mu}=1}$ & $\ensuremath{\underline{\kappa}=k},\ensuremath{\underline{\mu}=1}$\tabularnewline
\hline 
\multirow{2}{*}{Nakagami-$q$ (Hoyt)~\cite{moreno2016kappa}} & $\ensuremath{\underline{m_{t}}\rightarrow\infty},\ensuremath{\underline{m_{d}}=0.5},$ & \multirow{2}{*}{-} & $\ensuremath{\underline{m_{s}}\rightarrow\infty},\ensuremath{\underline{m_{d}}=0.5},$\tabularnewline
 & $\ensuremath{\underline{\kappa}=\frac{(1-\mbox{\ensuremath{q^{2}}})}{2q^{2}}},\ensuremath{\underline{\mu}=1}$ &  & $\ensuremath{\underline{\kappa}=\frac{(1-\mbox{\ensuremath{q^{2}}})}{2q^{2}}},\ensuremath{\underline{\mu}=1}$\tabularnewline
\hline 
\multirow{2}{*}{Nakagami-$m$ } & $\ensuremath{\underline{m_{t}}\rightarrow\infty},\ensuremath{\underline{m_{d}}\rightarrow\infty},$ & $\ensuremath{\underline{m_{t}}\rightarrow\infty},\ensuremath{\underline{m_{d}}\rightarrow\infty},$ & $\ensuremath{\underline{m_{s}}\rightarrow\infty},\ensuremath{\underline{m_{d}}\rightarrow\infty},$\tabularnewline
 & $\ensuremath{\underline{\kappa}\rightarrow0},\ensuremath{\underline{\mu}=m}$ & $\ensuremath{\underline{\kappa}\rightarrow0},\ensuremath{\underline{\mu}=m}$ & $\ensuremath{\underline{\kappa}\rightarrow0},\ensuremath{\underline{\mu}=m}$\tabularnewline
\hline 
\multirow{2}{*}{Rayleigh} & $\ensuremath{\underline{m_{t}}\rightarrow\infty},\ensuremath{\underline{m_{d}}\rightarrow\infty},$ & $\ensuremath{\underline{m_{t}}\rightarrow\infty},\ensuremath{\underline{m_{d}}\rightarrow\infty},$ & $\ensuremath{\underline{m_{s}}\rightarrow\infty},\ensuremath{\underline{m_{d}}\rightarrow\infty},$\tabularnewline
 & $\ensuremath{\underline{\kappa}\rightarrow0},\ensuremath{\underline{\mu}=1}$ & $\ensuremath{\underline{\kappa}\rightarrow0},\ensuremath{\underline{\mu}=1}$ & $\ensuremath{\underline{\kappa}\rightarrow0},\ensuremath{\underline{\mu}=1}$\tabularnewline
\hline 
\multirow{2}{*}{One-sided Gaussian} & $\ensuremath{\underline{m_{t}}\rightarrow\infty},\ensuremath{\underline{m_{d}}\rightarrow\infty},$ & $\ensuremath{\underline{m_{t}}\rightarrow\infty},\ensuremath{\underline{m_{d}}\rightarrow\infty},$ & $\ensuremath{\underline{m_{s}}\rightarrow\infty},\ensuremath{\underline{m_{d}}\rightarrow\infty},$\tabularnewline
 & $\ensuremath{\underline{\kappa}\rightarrow0},\ensuremath{\underline{\mu}=0.5}$ & $\ensuremath{\underline{\kappa}\rightarrow0},\ensuremath{\underline{\mu}=0.5}$ & $\ensuremath{\underline{\kappa}\rightarrow0},\ensuremath{\underline{\mu}=0.5}$\tabularnewline
\hline 
\end{tabular}
		\end{threeparttable}
	\end{table*}
%~\cite{newTCOM}, \cite{carlosratio}
	\begin{table*} 
	\small
	\renewcommand{\arraystretch}{1.5}
	  \captionsetup{justification=centering}
	\caption{Special Cases of the Double Shadowed $\kappa$-$\mu$ Type~II~(example~2), Type~III~(example~1) and Type~III~(example~2) Fading Models}
	\label{Table:2}
	\centering
	\begin{threeparttable}
	\begin{tabular}{|c|c|c|c|}
\hline 
\multirow{2}{*}{Fading models} & double shadowed $\kappa$-$\mu$ & double shadowed $\kappa$-$\mu$ & double shadowed $\kappa$-$\mu$  \tabularnewline
 & Type~II~(example~2) & Type~III~( example~1) & Type~III~(example~2) \tabularnewline
\hline 
\hline 
 Single shadowed $\kappa$-$\mu$ & \multirow{2}{*}{-} & \multirow{2}{*}{-} & \multirow{2}{*}{-}\tabularnewline
Type~I~(example~1)~\cite{6594884} &  &  & \tabularnewline
\hline 
 Single shadowed & $\ensuremath{\ensuremath{\underline{m_{s}}\rightarrow\infty},\ensuremath{\underline{m_{d}}=m_{d}},}$ & \multirow{2}{*}{-} & \multirow{2}{*}{-}\tabularnewline
$\kappa$-$\mu$ Type~I~(example~2) & $\ensuremath{\ensuremath{\underline{\kappa}=\kappa},\ensuremath{\underline{\mu}=\mu}}$ &  & \tabularnewline
\hline 
 Single shadowed  & $\ensuremath{\ensuremath{\underline{m_{s}}\rightarrow m_{s}},\ensuremath{\underline{m_{d}}\rightarrow}\infty,}$ & \multirow{2}{*}{-} & $\ensuremath{\ensuremath{\underline{m_{s}}\rightarrow m_{s}},\ensuremath{\underline{m_{t}}\rightarrow}\infty,}$\tabularnewline
$\kappa$-$\mu$ Type~II~(example~1) & $\ensuremath{\ensuremath{\underline{\kappa}=\kappa},\ensuremath{\underline{\mu}=\mu}}$ &  & $\ensuremath{\ensuremath{\underline{\kappa}=\kappa},\ensuremath{\underline{\mu}=\mu}}$\tabularnewline
\hline 
 Single shadowed & \multirow{2}{*}{-} & $\ensuremath{\ensuremath{\underline{m_{s}}\rightarrow m_{s}},\ensuremath{\underline{m_{t}}\rightarrow\infty},}$ & \multirow{2}{*}{-}\tabularnewline
$\kappa$-$\mu$ Type~II~(example~2) &  & $\ensuremath{\ensuremath{\underline{\kappa}=\kappa},\ensuremath{\underline{\mu}=\mu}}$ & \tabularnewline
\hline 
 Single shadowed $\kappa$-$\mu$  & \multirow{2}{*}{-} & $\ensuremath{\ensuremath{\underline{m_{s}}\rightarrow\infty},\ensuremath{\underline{m_{t}}\rightarrow m_{t}},}$ & \multirow{2}{*}{-}\tabularnewline
Type~III~(example~1)~\cite{carlos2017} &  & $\ensuremath{\ensuremath{\underline{\kappa}=\kappa},\ensuremath{\underline{\mu}=\mu}}$ & \tabularnewline
\hline 
Single shadowed $\kappa$-$\mu$ & \multirow{2}{*}{-} & \multirow{2}{*}{-} & $\ensuremath{\underline{m_{s}}=\infty},\ensuremath{\underline{m_{t}}\rightarrow}m_{t},$\tabularnewline
Type~III~(example~2) &  &  & $\ensuremath{\underline{\kappa}=\kappa},\ensuremath{\underline{\mu}=\mu}$\tabularnewline
\hline 
\multirow{2}{*}{$\eta$-$\mu$/inverse gamma} & \multirow{2}{*}{-} & \multirow{2}{*}{-} & \multirow{2}{*}{-}\tabularnewline
 &  &  & \tabularnewline
\hline 
\multirow{2}{*}{$\kappa$-$\mu$} & $\ensuremath{\underline{m_{s}}\rightarrow\infty},\underline{m_{d}}\rightarrow\infty,$ & $\ensuremath{\underline{m_{s}}\rightarrow\infty},\ensuremath{\underline{m_{t}}\rightarrow\infty},$ & $\ensuremath{\underline{m_{s}}\rightarrow\infty},\ensuremath{\underline{m_{t}}\rightarrow\infty},$\tabularnewline
 & $\ensuremath{\underline{\kappa}=\kappa},\ensuremath{\underline{\mu}=\mu}$ & $\ensuremath{\underline{\kappa}=\kappa},\ensuremath{\underline{\mu}=\mu}$ & $\ensuremath{\underline{\kappa}=\kappa},\ensuremath{\underline{\mu}=\mu}$\tabularnewline
\hline 
\multirow{2}{*}{$\eta$-$\mu$} & \multirow{2}{*}{-} & \multirow{2}{*}{-} & \multirow{2}{*}{-}\tabularnewline
 &  &  & \tabularnewline
\hline 
\color[rgb]{0,0,0}\multirow{2}{*}{Shadowed Rician~\cite{abdi2003new} } & \multirow{2}{*}{-} & \multirow{2}{*}{-} & \multirow{2}{*}{-}\tabularnewline
 &  &  & \tabularnewline
\hline 
\multirow{2}{*}{Rician} & $\ensuremath{\underline{m_{s}}\rightarrow\infty},\ensuremath{\underline{m_{d}}\rightarrow\infty},$ & $\ensuremath{\underline{m_{s}}\rightarrow\infty},\ensuremath{\underline{m_{t}}\rightarrow\infty},$ & $\ensuremath{\underline{m_{s}}\rightarrow\infty},\ensuremath{\underline{m_{t}}\rightarrow\infty},$\tabularnewline
 & $\ensuremath{\underline{\kappa}=k},\ensuremath{\underline{\mu}=1}$ & $\ensuremath{\underline{\kappa}= k},\ensuremath{\underline{\mu}=1}$ & $\ensuremath{\underline{\kappa}= k},\ensuremath{\underline{\mu}=1}$\tabularnewline
\hline 
\multirow{2}{*}{Nakagami-$q$ (Hoyt)~\cite{moreno2016kappa}} & \multirow{2}{*}{-} & \multirow{2}{*}{-} & \multirow{2}{*}{-}\tabularnewline
 &  &  & \tabularnewline
\hline 
\multirow{2}{*}{Nakagami-$m$ } & $\ensuremath{\underline{m_{s}}\rightarrow\infty},\ensuremath{\underline{m_{d}}\rightarrow\infty},$ & $\ensuremath{\underline{m_{s}}\rightarrow\infty},\ensuremath{\underline{m_{t}}\rightarrow\infty},$ & $\ensuremath{\underline{m_{s}}\rightarrow\infty},\ensuremath{\underline{m_{t}}\rightarrow\infty},$\tabularnewline
 & $\ensuremath{\underline{\kappa}\rightarrow0},\ensuremath{\underline{\mu}=m}$ & $\ensuremath{\underline{\kappa}\rightarrow0},\ensuremath{\underline{\mu}=m}$ & $\ensuremath{\underline{\kappa}\rightarrow0},\ensuremath{\underline{\mu}=m}$\tabularnewline
\hline 
\multirow{2}{*}{Rayleigh} & $\ensuremath{\underline{m_{s}}\rightarrow\infty},\ensuremath{\underline{m_{d}}\rightarrow\infty},$ & $\ensuremath{\underline{m_{s}}\rightarrow\infty},\ensuremath{\underline{m_{t}}\rightarrow\infty},$ & $\ensuremath{\underline{m_{s}}\rightarrow\infty},\ensuremath{\underline{m_{t}}\rightarrow\infty},$\tabularnewline
 & $\ensuremath{\underline{\kappa}\rightarrow0},\ensuremath{\underline{\mu}=1}$ & $\ensuremath{\underline{\kappa}\rightarrow0},\ensuremath{\underline{\mu}=1}$ & $\ensuremath{\underline{\kappa}\rightarrow0},\ensuremath{\underline{\mu}=1}$\tabularnewline
\hline 
\multirow{2}{*}{One-sided Gaussian} & $\ensuremath{\underline{m_{s}}\rightarrow\infty},\ensuremath{\underline{m_{d}}\rightarrow\infty},$ & $\ensuremath{\underline{m_{s}}\rightarrow\infty},\ensuremath{\underline{m_{t}}\rightarrow\infty},$ & $\ensuremath{\underline{m_{s}}\rightarrow\infty},\ensuremath{\underline{m_{t}}\rightarrow\infty},$\tabularnewline
 & $\ensuremath{\underline{\kappa}\rightarrow0},\ensuremath{\underline{\mu}=0.5}$ & $\ensuremath{\underline{\kappa}\rightarrow0},\ensuremath{\underline{\mu}=0.5}$ & $\ensuremath{\underline{\kappa}\rightarrow0},\ensuremath{\underline{\mu}=0.5}$\tabularnewline
\hline 
\end{tabular}
	\end{threeparttable}
\end{table*}

\section{Special Cases of the Example Double Shadowed $\kappa$-$\mu$ Fading Models}
%%%%%%%%%%%%%%%%%%%%%%%%%%%%%%%%%%%%%%%%%%%%%%%%%%%%%%%%%%%
%In this section, we present some special cases of the $\kappa$-$\mu$ double shadowed fading models proposed in this paper. As well as this, we demonstrate the usefulness of the $\kappa$-$\mu$ double shadowed Type~I fading model for directly characterizing the measured shadowed fading observed in D2D channels at 868~MHz.

\begin{figure}[t]
	\centering
	 \includegraphics[clip, trim={0.6cm 0.3cm 1.9cm 1.1cm},scale = 0.55]{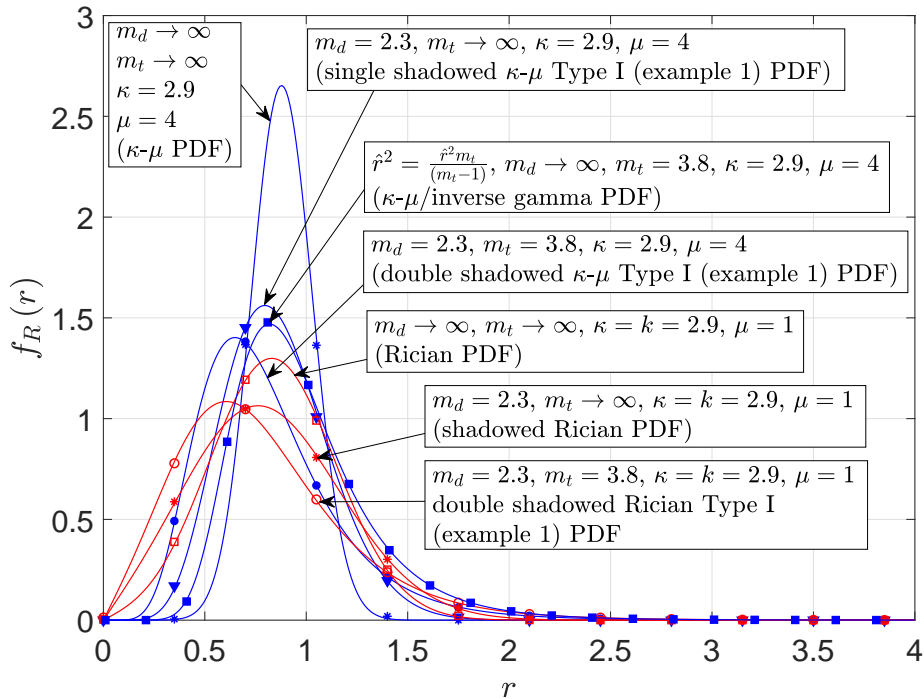}
	\caption{The PDF of the double shadowed $\kappa$-$\mu$ Type~I~(example~1) model reduced to some of its special cases: $\kappa$-$\mu$ (blue asterisk markers), single shadowed $\kappa$-$\mu$ Type~I~(example~1) (blue triangle markers), $\kappa$-$\mu$/inverse gamma (blue square markers), Rician (red square markers), shadowed Rician (red asterisk markers), double shadowed Rician Type~I~(example~1) (red circle markers). Here, $\hat{r} = 0.8$, lines represent analytical results, and the markers represent simulation results. 
} 
	\label{fig:img-6new}
\end{figure}

The PDFs given in~\eqref{eq_section3_02}, \eqref{eq_section3_10}, \eqref{eq_section3_08}, \eqref{eq_section3_14}, \eqref{eq_section3_12} and \eqref{eq_section3_13} represent an extremely versatile set of fading models as they inherit the generalities of the various types of single shadowed $\kappa$-$\mu$ fading model. It is recalled that in the double shadowed $\kappa$-$\mu$ Type~I model the $m_d$ parameter denotes the intensity of shadowing that the dominant signal component undergoes, whilst the $m_t$ parameter represents the degree of fluctuations that both the dominant and scattered signal components undergo as a result of the secondary shadowing process. 
Now, letting $m_{t} \rightarrow \infty$ in \eqref{eq_section3_02}, we obtain the PDF of the single shadowed $\kappa$-$\mu$ Type~I (example~1) model, whilst letting $m_{d} \rightarrow \infty$, we obtain the PDF of the single shadowed $\kappa$-$\mu$ Type~III (example~1) fading model. Allowing, $m_{d} \rightarrow \infty$ and $\hat{r}^{2} = \frac{m_{t}\hat{r}^{2}}{(m_{t}-1)}$ yields the $\kappa$-$\mu$/inverse gamma fading model. Hence, letting $m_{t} \rightarrow \infty$ and $m_{d} \rightarrow \infty$, we obtain the PDF of the $\kappa$-$\mu$ fading model. 
These special case results are illustrated in \figref{fig:img-6new} and are in exact agreement with the Monte Carlo (MC) simulations. The PDF of the $\eta$-$\mu$/inverse gamma fading model can also be obtained from the double shadowed $\kappa$-$\mu$ Type~I (example~1) fading model by setting $m_{d} \rightarrow \mu$, $\kappa=\frac{(1-\mbox{\ensuremath{\eta}})}{2\eta}$, $\mu=2\mu$ and $\hat{r}^{2} = \frac{m_{t}\hat{r}^{2}}{(m_{t}-1)}$. Thus, letting  $m_{t} \rightarrow \infty$, $m_{d} \rightarrow \mu$, $\kappa=\frac{(1-\mbox{\ensuremath{\eta}})}{2\eta}$ and $\mu=2\mu$ we obtain the PDF of the $\eta$-$\mu$ fading model. Likewise, the PDFs of the double shadowed Rician Type~I (example~1), shadowed Rician, and Rician fading models can be obtained from~\eqref{eq_section3_02} by first setting $\mu = 1$, $\kappa$ = $k$ (the Rician $k$-factor), followed by appropriate substitutions for $m_d$ and $m_t$.~\figref{fig:img-6new} shows the shape of the PDF for these special cases which are indicated in red. 

In a similar manner, the double shadowed $\kappa$-$\mu$ Type~I~(example~2) fading model contains the single shadowed $\kappa$-$\mu$ Type~I~(example~2) and Type~III~(example~1) fading models as special cases. Now letting $m_{t} \rightarrow \infty$ in \eqref{eq_section3_10}, we obtain the PDF of the single shadowed $\kappa$-$\mu$ Type~I~(example~2) model, whilst letting $m_{d} \rightarrow \infty$ we obtain the PDF of the single shadowed $\kappa$-$\mu$ Type~III~(example~1) fading model. Allowing both $m_{t} \rightarrow \infty$ and $m_{d} \rightarrow \infty$, the PDF of the $\kappa$-$\mu$ fading model is deduced. 
The PDF given in~\eqref{eq_section3_08} (double shadowed Type~II~(example~1)) also represents an extremely flexible fading model as it contains the single shadowed $\kappa$-$\mu$ Type~I~(example~1), Type~II~(example~2), $\kappa$-$\mu$ and $\eta$-$\mu$ fading models as special cases. Different from the double shadowed $\kappa$-$\mu$ Type~I~(example~1) and Type~I~(example~2) fading models, here the $m_s$ parameter represents the degree of fluctuation that the scattered signal components undergo. Now, letting $m_{s} \rightarrow \infty$ in~\eqref{eq_section3_08}, we obtain the PDF of the single shadowed $\kappa$-$\mu$ Type~I~(example~1) model, whilst letting $m_{d} \rightarrow \infty$ we obtain the PDF of the single shadowed $\kappa$-$\mu$ Type~II (example~2) fading model. Allowing both $m_{s} \rightarrow \infty$ and $m_{d} \rightarrow \infty$ in~\eqref{eq_section3_08}, the double shadowed $\kappa$-$\mu$ Type~II~(example~1) fading model coincides with the $\kappa$-$\mu$ fading model. For the reader's convenience, Table~III summarizes the special cases of the double shadowed $\kappa$-$\mu$~Type~I~(example~1), Type~I~(example~2) and Type~II~(example~1) fading models whilst Table~IV summarizes the special cases of the double shadowed $\kappa$-$\mu$~Type~II (example~2), Type~III~(example~1) and Type~III~(example~2) fading models. For the sake of clarity, the double shadowed $\kappa$-$\mu$ parameters have been underlined. 

\section{Numerical Results and Channel Measurements}
%%%%%%%%%%%%%%%%%%%%%%%%%%%%%%%%%%%%%%%%%%%%%%%%%%%%%%%%%%%
%%%%%%%%%%%%%%%%%%%%%%%%%%%%%%%%%%%%%%%%%%%%%%%%%%%%%%%%%%%
\subsection{Numerical Results}
%%%%%%%%%%%%%%%%%%%%%%%%%%%%%%%%%%%%%%%%%%%%%%%%%%%%%%%%%%%
\begin{figure}[t]
	\centering
	 \includegraphics[clip, trim={0.6cm 0.3cm 1.9cm 1.1cm},scale = 0.45]{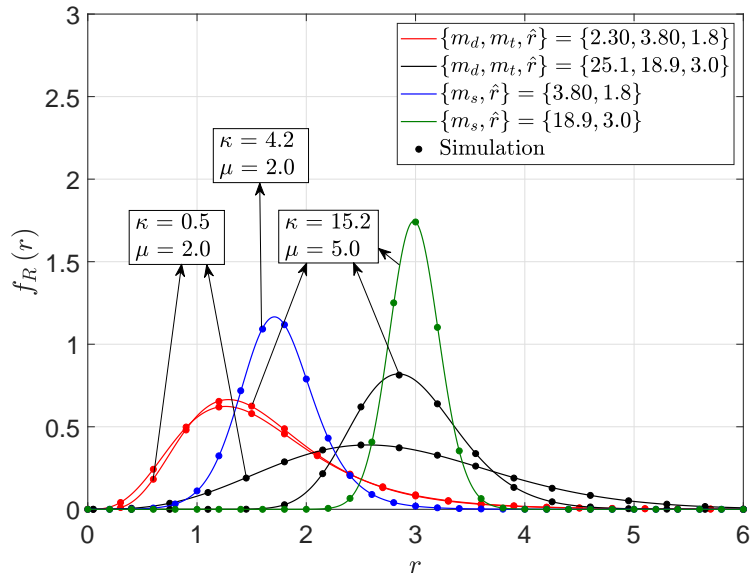}
	\caption{The PDF of the single shadowed $\kappa$-$\mu$ Type~II~(example~2) (blue and green lines) and double shadowed $\kappa$-$\mu$ Type~I~(example~1) (red and black lines) fading models. Lines represent the analytical results, circle markers represent simulation results. 
} 
	\label{fig:img-1}
\end{figure}
\begin{figure}[t]
	\centering
	 \includegraphics[clip, trim={0.6cm 0.3cm 1.9cm 1.1cm},scale = 0.45]{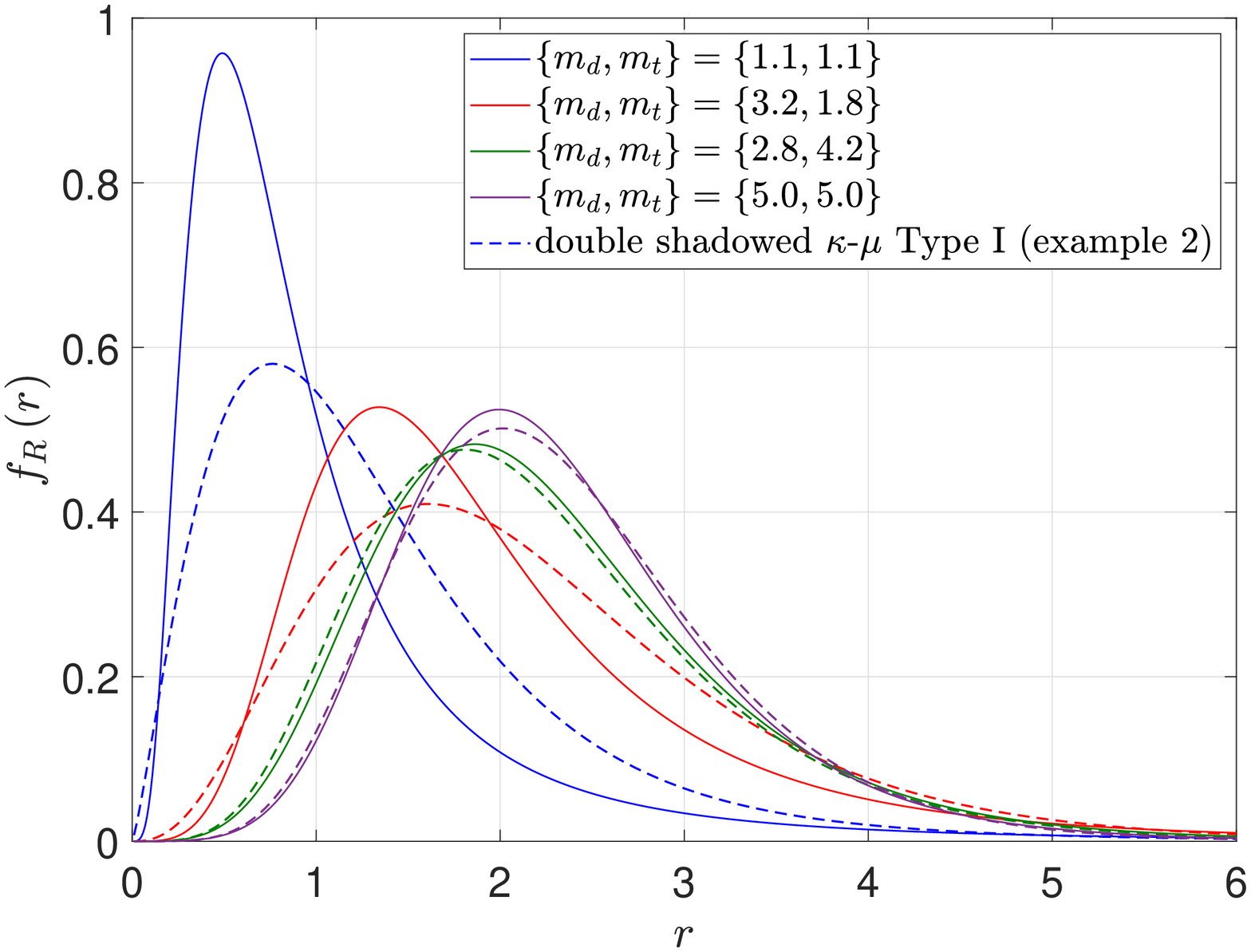}
\caption{The PDF of the double shadowed $\kappa$-$\mu$ Type~I~(example~1) and (example~2) fading models for different values of $m_d$ and $m_t$. Here, $\kappa$ = 3.9, $\mu$~=~2.4, and $\hat{r} = 2.5$. Solid lines represent the PDF of the double shadowed $\kappa$-$\mu$ Type~I~(example~1) model and dashed lines represent the PDF of the double shadowed $\kappa$-$\mu$ Type~I~(example~2) model. } 
	\label{fig:img-2}
\end{figure}

Figs.~\ref{fig:img-1} and \ref{fig:img-2} show some plots of the PDF of the single shadowed $\kappa$-$\mu$ Type~II~(example~2), and double shadowed Type~I~(example~1) models for different values of $\kappa$, $\mu$, $m_{s}$, $m_{d}$, $m_{t}$, and $\hat{r}$. It is noted that the values of the parameters are chosen to illustrate the wide range of shapes that the new shadowed fading models can exhibit. \figref{fig:img-1} shows the PDF of the single shadowed $\kappa$-$\mu$ Type~II~(example~2) and double shadowed Type~I~(example~1) fading models for $\{\kappa,\mu\}=\{0.5,2.0\}, \{4.2,2.0\},\{15.1,5.0\}$, $\{m_{d},m_{t},\hat{r}\}=\{2.3,3.8,1.8\},\{25.1,18.9,3.0\}$ and $\{m_{s},\hat{r}\}=\{3.8,1.8\},\{18.9,3.0\}$. In all cases, the analytical results agree with the MC simulations, which verifies their validity. 

%%%%%%%%%%%%%%%%%%%%%%%%%%%%%%%%%%%%%%%%%%%%%%%%%%%%%%%%%%%%%%%%%%%%%%%%%%%%%%%%%%
\subsection{Channel Measurements}
%%%%%%%%%%%%%%%%%%%%%%%%%%%%%%%%%%%%%%%%%%%%%%%%%%%%%%%%%%%%%%%%%%%%%%%%%%%%%%%%%%

While a detailed empirical investigation of all of the models presented here is clearly beyond the scope and space constraints of the paper, in this subsection we take the double shadowed $\kappa$-$\mu$ Type~I as an example and compare it with some BAN channel measurements which were conducted at 2.45~GHz. 

\begin{figure}[t]
	\centering
\includegraphics[clip, trim={0.3cm 0.3cm 0.3cm 0.2cm},scale = 0.35]{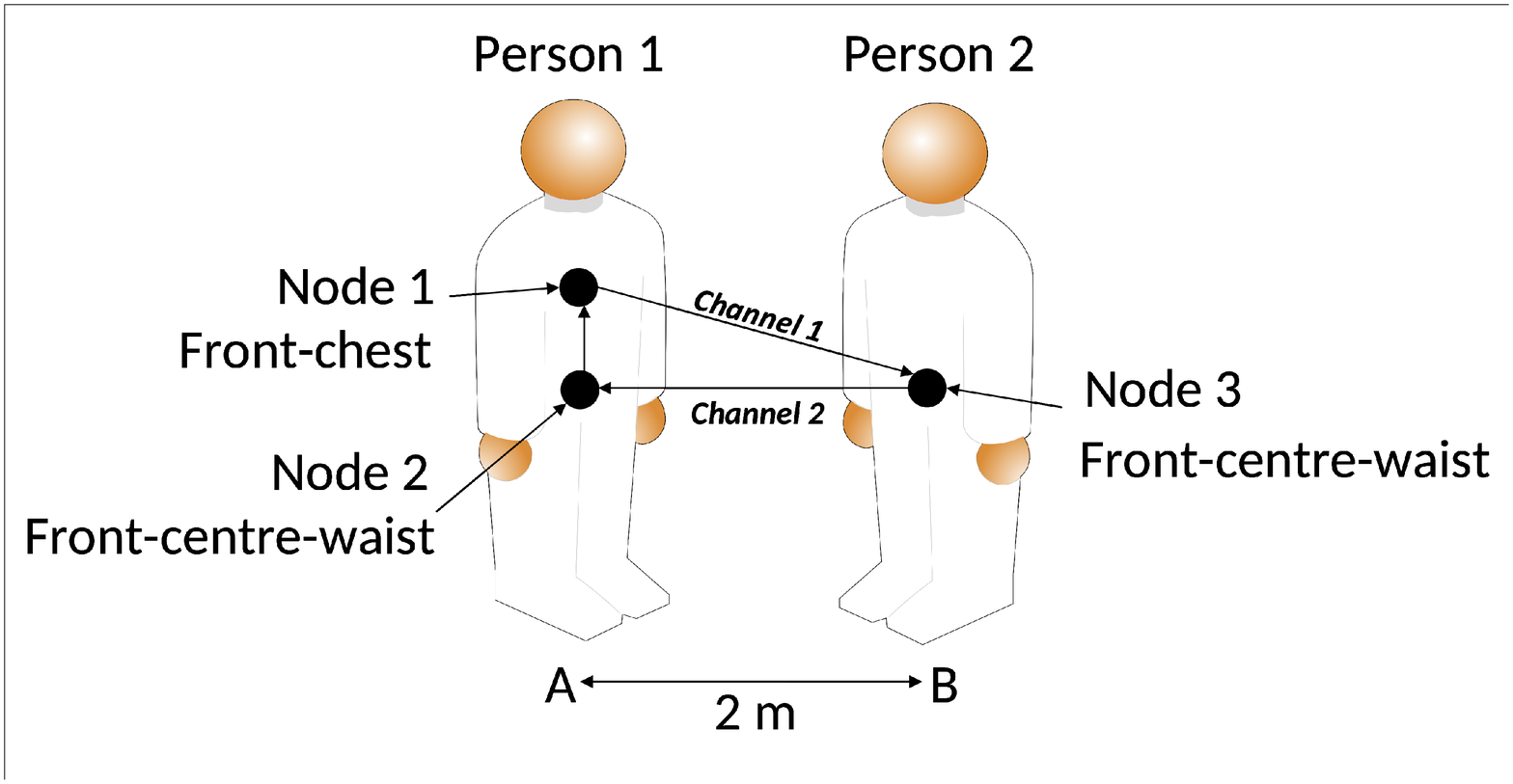}
	\caption{On-body transceivers on front-chest of person 1 (node 1), front-centre-waist of person 1 (node 2) and front-centre-waist of person 2 (node 3). Channel 1 represents the channel between nodes 1 and 3 and channel 2 represents the channel between nodes 3 and 2.}
	\label{fig:img-30}
\end{figure}

The channel measurements considered a three node BAN system (see Fig.~\ref{fig:img-30}) with a sample frequency of 198.4~Hz. Nodes~1 and~2 were positioned on the front-centre-waist and front-chest of person~1, whilst BAN node~3 was placed on the front-centre-waist of person~2. Note that the BAN-to-BAN links which exist between nodes~1 and 3, and nodes~3 and 2 are referred to as channel~1 and channel~2, respectively.
The users were initially instructed to stand motionless at positions A and B, separated by a distance of 2~m. Both test subjects were then instructed to walk around randomly within a circle of radius 0.5~m from their starting positions. This is representative of a situation which may lead to the propagation characteristics described by the double shadowed $\kappa$-$\mu$ Type~I model. In this case, the signal power delivered through the optical path between the transmitter and receiver will be shadowed by objects moving within its locality (e.g., blockages to the dominant component due to the random movements performed by person 1 and person 2), whilst further shadowing of the received power (combined multipath and dominant paths) may also occur due to obstacles moving in the vicinity of the transmitter or receiver (e.g., blockages caused by person 1 and 2 moving). 
The measurements were performed in a reverberation chamber with the mechanical stirrers disabled, which facilitated a study of the envelope variation solely due to body movements. A more detailed description of the measurement environment (reverberation chamber), the experimental set up and scenario can be found in~\cite{bhargav2016simultaneous} and~\cite{bhargav2016experimental}.

\begin{table}[t] 
\small
\renewcommand{\arraystretch}{1.4}
  \captionsetup{justification=centering}
\caption{Estimated parameters for Double shadowed $\kappa$-$\mu$ Type I (example~1), $\kappa$-$\mu$, Rice and Nakagami-$m$ fading models for BAN scenario along with the computed AIC scores and ranks.}
\label{Table:7}
\centering
\begin{threeparttable}
\begin{tabular}{|>{\centering}p{2cm}|>{\centering}p{2.5cm}|>{\centering}p{2.5cm}|>{\centering}p{2.5cm}|>{\centering}p{2.5cm}|}
\hline 
Fading model & Parameters for Channel 1  & AIC scores and ranks for Channel 1 & Parameters for Channel 2  & AIC scores and ranks for Channel 2\tabularnewline
\hline 
\hline
Double shadowed $\kappa$-$\mu$ Type I (example 1) & $\kappa=45.17$; $\mu=0.54$; $\hat{r}=1.27$; $m_{d}=1.31$; $m_{t}=2.24$ & 7366.5

(Rank 1) & $\kappa=5.91$; $\mu=0.79$; $\hat{r}=1.26$; $m_{d}=1.01$; $m_{t}=2.99$ & 6607.2

(Rank 1)\tabularnewline
\hline 
$\kappa$-$\mu$ & $\kappa=0.10$; $\mu=0.92$; $\hat{r}=1.07$; & 7571.5

(Rank 4) & $\kappa=0.40$; $\mu=0.77$; $\hat{r}=1.10$; & 6736.8

(Rank 3)\tabularnewline
\hline 
Rice & $k=0.001$

$\hat{r}=1.16$; & 7509.4

(Rank 3) & $k=0.001$

$\hat{r}=1.18$; & 6836.0

(Rank 4)\tabularnewline
\hline 
Nakagami-$m$ & $m=0.84$;

$\hat{r}=1.16$; & 7416.2

(Rank 2) & $m=0.76$;

$\hat{r}=1.18$; & 6614.5

(Rank 2)\tabularnewline
\hline 
\end{tabular}
\end{threeparttable}
\end{table}

\begin{figure}[t]
	\centering
	 \includegraphics[clip, trim={0.6cm 0.1cm 1.9cm 0.2cm},scale = 0.44]{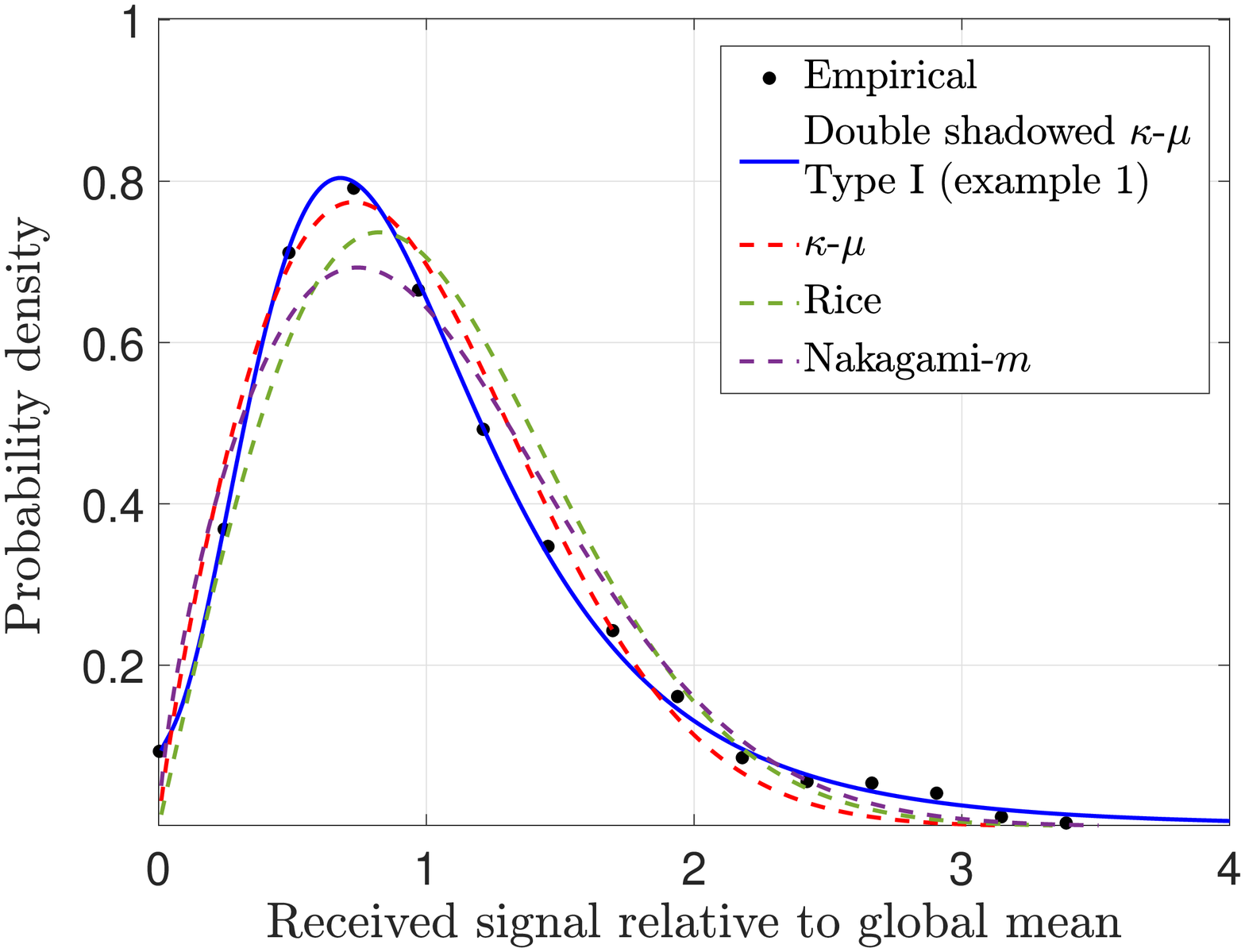}
	\caption{Empirical (markers) and theoretical (lines) probability densities for BAN Channel~1 measurements.} 
	\label{fig:img-50}
\end{figure}
\begin{figure}[t]
	\centering
	 \includegraphics[clip, trim={0.6cm 0.2cm 1.9cm 0.2cm},scale = 0.44]{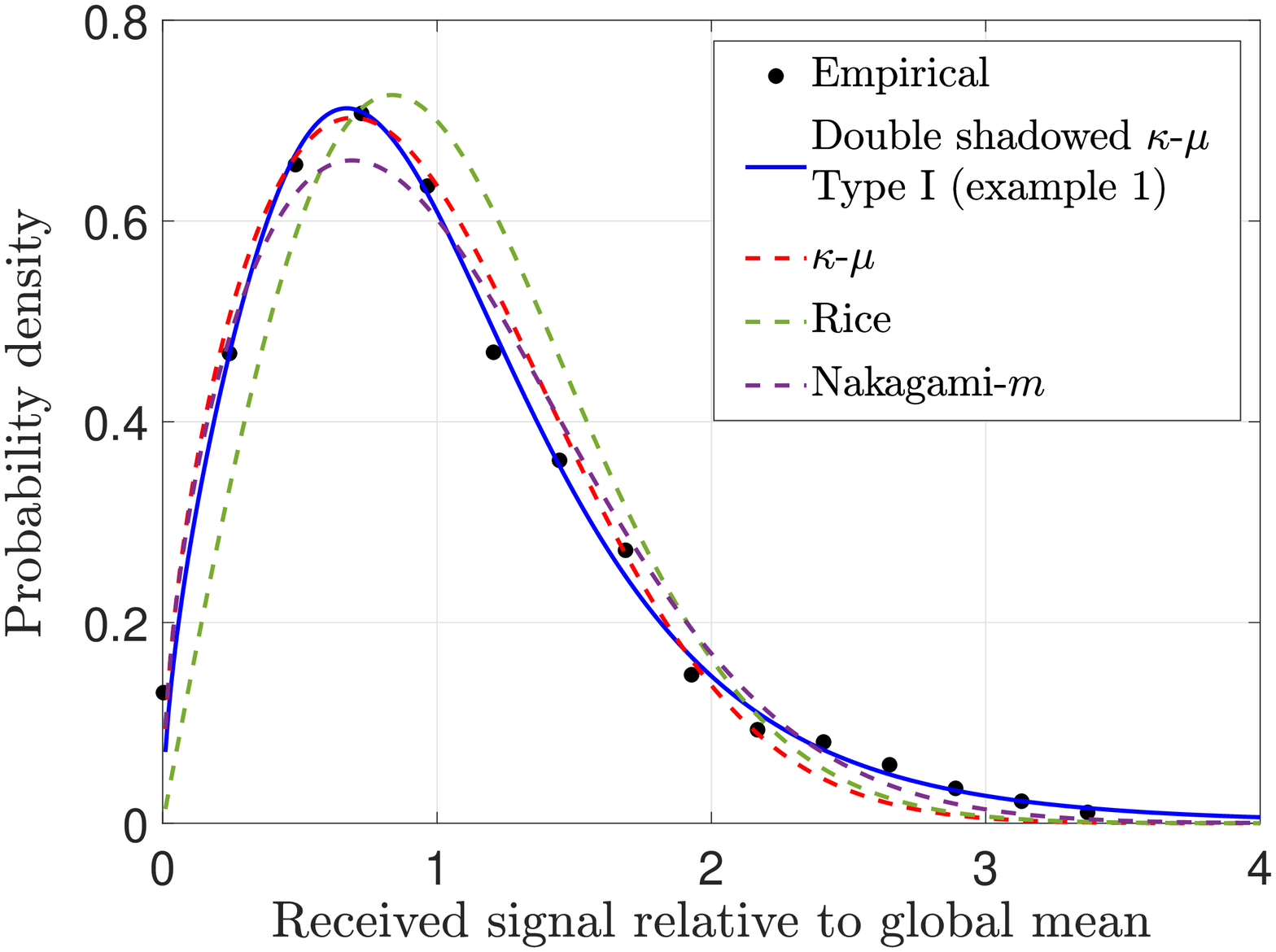}
	\caption{Empirical (markers) and theoretical (lines) probability densities for BAN Channel~2 measurements.} 
	\label{fig:img-51}
	\end{figure}

For the analysis, the global mean signal power was removed from the BAN measurement data. As an example of the data fitting process, Figs.~\ref{fig:img-50} and~\ref{fig:img-51} show the PDF of the double shadowed $\kappa$-$\mu$ Type~I (example 1) fading model fitted to the BAN data for channels~1 and 2, respectively. The figures also compare the empirical data with popular fading models such as $\kappa$-$\mu$, Rice and Nakagami-$m$. It can clearly be seen that the double shadowed $\kappa$-$\mu$ Type~I (example 1) fading model provides a superior fit to the experimental data when compared to $\kappa$-$\mu$, Rice and Nakagami-$m$. 
 
All parameter estimates were obtained using the \texttt{lsqnonlin} function available in the optimization toolbox of MATLAB along with either the double shadowed $\kappa$-$\mu$ Type~I (example 1), $\kappa$-$\mu$, Rice and Nakagami-$m$ PDFs. To allow the reader to reproduce these plots, the parameter estimates obtained are provided in Table~V. Referring to the parameter estimates obtained for double shadowed $\kappa$-$\mu$ Type~I (example 1) model, we observe high values of $\kappa$ indicating that these channels are composed of strong dominant signal components. As well as this, it is seen that the dominant component ($m_d$) experienced significant shadowing caused by person 1 and person 2 exhibiting random movements (illustrated by low estimates of $m_d$ observed). Further shadowing of the received power (combined multipath and dominant paths) is also indicated through the low values of $m_t$ obtained. This was again caused by both persons (with the on-body nodes) performing random movements. Recall that the mechanical stirrers in the reverberation chamber were disabled. Therefore, in this instance, the chamber represents a highly reflective environment with a finite number of multipath components. This is illustrated by the low estimates of $\mu$ observed~\cite{5997282}.

Table~V also presents the computed Akaike Information Criterion (AIC). The AIC is an important metric, which not only considers the fit but also penalizes model complexity. From Table~V, we can clearly see that for each of the considered BAN channels, the double shadowed $\kappa$-$\mu$ Type~I (example~1) model outperformed the conventional fading models considered here. More importantly though, it reports that double shadowing exists for these scenarios, something that cannot be characterized using the existing fading models.

\section{Conclusion}
%%%%%%%%%%%%%%%%%%%%%%%%%%%%%%%%%%%%%%%%%%%%%%%%%%%%%%%%%%%
For the first time in the literature, this paper has discussed the various ways in which a $\kappa$-$\mu$ fading envelope can be affected by shadowing. A family of shadowed $\kappa$-$\mu$ fading models were proposed and classified based on whether the underlying $\kappa$-$\mu$ envelope undergoes single or double shadowing. 
%Special conditions of both classes of the $\kappa$-$\mu$ shadowed fading models were described, differentiated by the way in which the dominant components, scattered waves or both are fluctuated due to the shadowing phenomenon. 
In total, three types of single shadowed $\kappa$-$\mu$ model (Type I - III)  were introduced. It is worth
emphasizing that these model frameworks are general and do not depend on predefined RVs that are responsible for shaping the shadowing characteristics. However, for illustrative purposes, two example cases for each type of single shadowed model were discussed where it was assumed that the shadowing is shaped by a Nakagami-$m$ RV or an inverse Nakagami-$m$ RV. A further three types of double shadowed $\kappa$-$\mu$  model (Type~I - III) were also introduced, all of which are novel. Similar to the single shadowed models, two example cases were also provided for each type of double shadowed model.
 %The classification of the single shadowed models was dependent on whether the underlying $\kappa$-$\mu$ fading assumed that the dominant component, scattered waves or both experienced shadowing; and whether this shadowing was influenced by a Nakagaim-$m$ or an inverse Nakagami-$m$ random process.
%It is also entirely possible that in addition to the dominant component or the scattered waves being shadowed, further shadowing may also occur which impacts the scattered signal, and also administers secondary shadowing to the already perturbed dominant component.  
%These models were classified depending on whether the underlying $\kappa$-$\mu$ fading assumed that in addition to shadowing of the dominant component or the scattered waves, a secondary round of shadowing also occurs; or whether the dominant component and scattered waves of a $\kappa$-$\mu$ fading envelope were subject to two independent shadowing processes; and whether these shadowing processes were shaped by a Nakagami-$m$ or an inverse Nakagami-$m$ RV. 
%The $\kappa$-$\mu$ double shadowed Type I model represents a $\kappa$-$\mu$ fading channel that undergoes LOS shadowing and then experiences additional multiplicative shadowing or vice versa. The Type II model describes a $\kappa$-$\mu$ fading scenario in which the shadowing of the dominant component is shaped by a Nakagami-$m$ RV and the scattered waves are subject to fluctuations influenced by an inverse Nakagami-$m$ random process. 
Finally, the utility of the double shadowed $\kappa$-$\mu$ Type~I model was illustrated for characterizing the shadowed fading encountered in BAN channels. It was shown that this model provides a superior fit to the channel data when compared to  popular fading models such as $\kappa$-$\mu$, Rice and Nakagami-$m$. Crucially though, this fading model was able to provide an excellent characterization of the field data without the need to determine a smoothing window size to abstract the local mean signal. 

%%%%%%%%%%%%%%%%%%%%%%%%%%%%%%%%%%%%%%%%%%%%%%%%%%%%%%%%%%%
	\appendices
	%%%%%%%%%%%%%%%%%%%%%%%%%%%%%%%%%%%%%%%%%%%%%%%%%%%%%%%%%%%
	
\appendix
\setcounter{secnumdepth}{0}
\section{Appendix}	%%%%%%%%%%%%%%%%%%%%%%%%%%%%%%%%%%%%%%%%%%%%%%%%%%%%%%%%%%%
	\subsection{A. PROOF OF THEOREM 1 $\label{app:A0}$}
	%%%%%%%%%%%%%%%%%%%%%%%%%%%%%%%%%%%%%%%%%%%%%%%%%%%%%%%%%%%
	Considering the signal model given in~\eqref{eq_section2_01} where $\xi$ is assumed to be an inverse Nakagami-$m$ RV with shape parameter $m_{d}$ and $\mathbb{E}\left[{\xi}^2 \right] = 1$, its PDF is given by
	\begin{equation}
f_{\xi}\left(\xi\right)=\frac{2(m_{d}-1)^{m_{d}}}{\Gamma\left(m_{d}\right)\xi^{2m_{d}+1}}\mathrm{e}^{-\frac{(m_{d}-1)}{\xi^{2}}}.
\label{eq_app_a2}
\end{equation}
	
	To determine the envelope distribution of the single shadowed $\kappa$-$\mu$ Type~I~(example~2) fading model we average the conditional PDF, $f_{R|\xi}\left(r|\xi \right)$, with the PDF of $\xi$ given in~\eqref{eq_app_a2} i.e.
	\begin{equation}
	f_{R}\left(r\right)=\intop_{0}^{\infty}f_{R|\xi}\left(r|\xi\right)f_{\xi}\left(\xi\right)d\xi. 
	\label{eq_app_a3}
	\end{equation}
	
	The signal model for the single shadowed $\kappa$-$\mu$ Type~I~(example~2) fading model, insinuates that the conditional probability, $f_{R|\xi}\left(r|\xi\right)$,  follows a $\kappa$-$\mu$ distribution with PDF~\cite{4231253}
	\begin{equation}
	f_{R|\xi}\left(r|\xi\right)=\frac{r^{\mu}}{\sigma^{2}\left(\xi d\right){}^{\mu-1}}\mathrm{e}^{\frac{-r^{2}-\xi^{2}d^{2}}{2\sigma^{2}}}\mathrm{I}_{\mu-1}\left(\frac{\xi dr}{\sigma^{2}}\right)
	\label{eq_app_a1new}
	\end{equation}
	where $d^2$ and $\sigma^2$ are as defined in section II.A (also see~\cite{4231253}). 
	
	An analytical expression for the PDF of the single shadowed $\kappa$-$\mu$ {Type~I~(example~2)} fading model can be obtained by substituting~\eqref{eq_app_a1new} and~\eqref{eq_app_a2} in~\eqref{eq_app_a3} as follows:
	
	\begin{equation}
	\!f_{R}\!\left(r\right)\!=\!\!\intop_{0}^{\infty}\frac{2r^{\mu}(m_{d}-1)^{m_{d}}~\mathrm{e}^{-\frac{(m_{d}-1)}{\xi^{2}}-\frac{r^{2}+\xi^{2} d^{2}}{2\sigma^{2}}}}{\sigma^{2}\left(\xi d\right)^{\mu-1}\Gamma\left(m_{d}\right)\xi^{2m_{d}+1}}~\mathrm{I}_{\mu-1}\!\!\left(\!\frac{\xi d r}{\sigma^{2}}\!\right)d\xi.
	\label{eq_app_a2new}
	\end{equation}
	Replacing the modified Bessel function of the first kind with~\cite[03.02.02.0001.01]{wolfram} i.e, 
	${I_v}(x) = \mathop \sum \limits_{k = 0}^\infty  \frac{{{{\left( {\frac{x}{2}} \right)}^{v + 2k}}}}{{k!{\rm{\Gamma }}\left( {v + k + 1} \right)}}$
	in~\eqref{eq_app_a2new}, followed by solving the integral using~\cite[eq. 3.471.9]{TofI}, and finally substituting $d = \sqrt{2 \mu {\sigma^{2}} \kappa}$ along with $\sigma=\sqrt{\frac{\hat{r}^{2}}{2\mu\left(1+\kappa\right)}}$ in the resultant expression, we obtain~\eqref{eq_section2_new1}.
	%%%%%%%%%%%%%%%%%%%%%%%%%%%%%%%%%%%%%%%%%%%%%%%%%%%%%%%%%%%
	\subsection{B. PROOF OF THEOREM~2 AND THEOREM~3 $\label{app:A}$}
	%%%%%%%%%%%%%%%%%%%%%%%%%%%%%%%%%%%%%%%%%%%%%%%%%%%%%%%%%%%

	Assuming that $\xi$ is a Nakagami-$m$ RV  with shape parameter $m_{s}$ and $\mathbb{E}\left[{\xi}^2 \right] = 1$, its PDF is given by
	
	\begin{equation}
	f_{\xi}\left(\xi\right)=\frac{2m_{s}^{m_{s}}\xi^{2m_{s}-1}}{\Gamma\left(m_{s}\right)}\mathrm{e}^{-{m_{s}}{\xi^{2}}}.
	\label{eq_app_a1}
	\end{equation}
		The signal model presented in~\eqref{eq_section2_03} insinuates that the conditional probability, $f_{R|\xi}\left(r|\xi\right)$, follows a $\kappa$-$\mu$ distribution with PDF~\cite{4231253}
	
	\begin{equation}
	f_{R|\xi}\left(r|\xi\right)=\frac{r^{\mu}}{\sigma^{2}\xi^{2}d^{\mu-1}}\mathrm{e}^{\frac{-r^{2}-d^{2}}{2\sigma^{2}\xi^{2}}}\mathrm{I}_{\mu-1}\left(\frac{dr}{\sigma^{2}\xi^{2}}\right).
	\label{eq_app_a4}
	\end{equation}
		An analytical expression for the PDF of the single shadowed $\kappa$-$\mu$ Type~II~(example~1) fading model can be obtained by substituting~\eqref{eq_app_a4} and~\eqref{eq_app_a1} in~\eqref{eq_app_a3} as follows:
	
	\begin{equation}
	\!f_{R}\left(r\right)=\intop_{0}^{\infty}\frac{2r^{\mu}m_{s}^{m_{s}}\xi^{2m_{s}-3}}{\sigma^{2}d^{\mu-1}\Gamma\left(m_{s}\right)}\mathrm{e}^{-m_{s}\xi^{2}-\frac{r^{2}+d^{2}}{2\sigma^{2}\xi^{2}}}~\mathrm{I}_{\mu-1}\!\!\left(\!\frac{dr}{\sigma^{2}\xi^{2}}\!\right)d\xi.
	\label{eq_app_a5}
	\end{equation}
		Replacing the modified Bessel function of the first kind with its series representation~\cite[03.02.02.0001.01]{wolfram} in~\eqref{eq_app_a5}, followed by solving the integral using~\cite[eq. 3.471.9]{TofI}, and finally substituting $d = \sqrt{2 \mu {\sigma^{2}} \kappa}$; $\sigma=\sqrt{\frac{\hat{r}^{2}}{2\mu\left(1+\kappa\right)}}$ in the resultant expression, we obtain~\eqref{eq_section2_04}.
	
	Similarly, a closed form expression for the PDF of the single shadowed $\kappa$-$\mu$ Type~II~(example~2) fading model, is obtained by substituting~\eqref{eq_app_a4} and~\eqref{eq_app_a2} (after replacing $m_d$ with $m_s$) in~\eqref{eq_app_a3}, which yields 
	
	\begin{equation}
	\!f_{R}\left(r\right)\!=\!\intop_{0}^{\infty}\frac{2r^{\mu}(m_{s}-1)^{m_{s}} \mathrm{e}^{-\frac{(m_{s}-1)}{\xi^{2}}-\frac{r^{2}+d^{2}}{2\sigma^{2}\xi^{2}}}}{\sigma^{2}d^{\mu-1}\Gamma\left(m_{s}\right) \xi^{2m_{s}+3}}~\mathrm{I}_{\mu-1}\!\!\left(\!\frac{dr}{\sigma^{2}\xi^{2}}\!\right)d\xi.
	\label{eq_app_a6}
	\end{equation}
The above integral is identical to~\cite[eq. 2.15.3.2]{prudinkov}. 
	Now, substituting for $d$ and $\sigma$ in the resultant expression and performing some algebraic manipulations, we obtain~\eqref{eq_section2_06}.

	\subsection{C. PROOF OF THEOREM~\ref{thm Type5pdf}$\label{app:C}$}
	%%%%%%%%%%%%%%%%%%%%%%%%%%%%%%%%%%%%%%%%%%%%%%%%%%%%%%%%%%%
	
	We determine the envelope distribution of the single shadowed $\kappa$-$\mu$ Type~III~(example~1) fading model using~\eqref{eq_app_a3}. Here, the conditional probability, $f_{R|\xi}\left(r|\xi\right)$ is given by
	
	\begin{equation}
	f_{R|\xi}\left(r|\xi\right)\!=\!\!\frac{2\mu\!\left(1\!+\!\kappa\right)^{\frac{\mu+1}{2}}\!r^{\mu}\mathrm{e}^{\frac{-\mu\left(1+\kappa\right)r^{2}}{\xi^{2}\hat{r}^{2}}}}{\kappa^{\frac{\mu-1}{2}}e^{\kappa\mu}\left(\xi^{2}\hat{r}^{2}\right)^{\frac{\mu+1}{2}}}\mathrm{I}_{\mu-1}\!\!\left(\!\frac{2\mu\sqrt{\kappa\left(1\!+\!\kappa\right)}r}{\sqrt{\xi^{2}\hat{r}^{2}}}\!\right). 
	\label{eq_app_c1}
	\end{equation}
	Substituting~\eqref{eq_app_c1} and~\eqref{eq_app_a1} (after replacing $m_s$ with $m_t$) in~\eqref{eq_app_a3}, followed by replacing the modified Bessel function of the first kind with its series representation~\cite[03.02.02.0001.01]{wolfram},
	and solving the resulting integral using~\cite[eq. 3.471.9]{TofI}, we obtain~\eqref{eq_section2_15}.
	
%%%%%%%%%%%%%%%%%%%%%%%%%%%%%%%%%%%%%%%%%%%%%%%%%%%%%%%%%%%
	\subsection{D. PROOF OF THEOREM~\ref{thm Type1apdf} $\label{app:D}$}
	%%%%%%%%%%%%%%%%%%%%%%%%%%%%%%%%%%%%%%%%%%%%%%%%%%%%%%%%%%%
	We determine the envelope distribution of the double shadowed $\kappa$-$\mu$ Type~I~(example~1) fading model using~\eqref{eq_app_a3} (after replacing $\xi$ with $\alpha$).  Its signal model insinuates that the conditional probability, $f_{R|\alpha}\left(r|\alpha\right)$, follows a single shadowed $\kappa$-$\mu$ Type~I~(example~1) distribution with PDF~\cite{6594884}~\cite{cotton2015human} 
	
	\begin{equation}
	f_{R|A}\left(r|\alpha\right)=\frac{2\mu^{\mu}\left(1+\kappa\right)^{\mu}r^{2\mu-1}m_{d}^{m_{d}}}{\Gamma\left(\mu\right)\left(m_{d}+\mu\kappa\right)^{m_{d}}\alpha^{2\mu}\hat{r}^{2\mu}} 
	\\ \mathrm{e}^{-\mu\left(1+\kappa\right)\frac{r^{2}}{\alpha^{2}\hat{r}^{2}}}{}_{1}F_{1}\left(m_{d};\mu;\frac{\mu^{2}\kappa\left(1+\kappa\right)r^{2}}{\alpha^{2}\hat{r}^{2}\left(m_{d}+\mu\kappa\right)}\right) \label{eq_app_d1}
	\end{equation}
	where, $\kappa$, $\mu$, $\hat{r}$ and $m_d$ are as defined in section II~A. 
		Now replacing $\xi$ with $\alpha$, and $m_d$ with $m_t$ in~\eqref{eq_app_a2}, followed by substituting the resultant expression and \eqref{eq_app_d1} in \eqref{eq_app_a3} (where $\xi$ is replaced with $\alpha$), we obtain
	
	\begin{equation}
\!\!f_{R}\left(r\right)\!  =\frac{4m_{d}^{m_{d}}(m_{t}-1)^{m_{t}}\mu^{\mu}\left(1+\kappa\right)^{\mu}r^{2\mu-1}}{\Gamma\left(\mu\right)\Gamma\left(m_{t}\right)\left(m_{d}+\mu\kappa\right)^{m_{d}}\hat{r}^{2\mu}}\!\!\intop_{0}^{\infty}\!\!\frac{\!\alpha^{-2m_{t}-2\mu-1}}{\mathrm{e}^{\left(\mu\left(1+\kappa\right)\frac{r^{2}}{\hat{r}^{2}}+(m_{t}-1)\right)\frac{1}{\alpha^{2}}}}
	{}_{1}F_{1}\!\left(\!m_{d};\mu;\frac{\mu^{2}\kappa\left(1\!+\!\kappa\right)r^{2}}{\alpha^{2}\hat{r}^{2}\left(m_{d}+\mu\kappa\right)}\right)d\alpha.
	\label{eq_app_d2}
	\end{equation}
		The above integral is identical to~\cite[eq. 7.621.4]{TofI}. Performing the necessary transformation of variables followed by some simple mathematical manipulations, we obtain~\eqref{eq_section3_02}.

	{\subsection{E. PROOF OF THEOREM~\ref{thm TypeIIIpdf}$\label{app:H}$}
		%%%%%%%%%%%%%%%%%%%%%%%%%%%%%%%%%%%%%%%%%%%%%%%%%%%%%%%%%%%	
		The signal envelope, $R$, of the double shadowed $\kappa$-$\mu$ Type~I~(example~2) fading model is given by \eqref{eq_section3_01}. Here, $A$ follows a Nakagami-$m$ distribution with shape parameter $m_t$, and $\xi$ follows an inverse Nakagami-$m$ distribution with shape parameter $m_d$. It is noted that the model in \eqref{eq_section3_01} may be viewed as a product of a Nakagami-$m$ RV and a single shadowed $\kappa$-$\mu$ Type~I~(example~2) RV. According to standard probability procedure, this PDF can be obtained as 
		
		\begin{equation}\label{eq_app_h1}
		f_R(r) = \int_0^\infty \frac{1}{a} f_T\left(\frac{r}{a}\right) f_A(a) \,da
		\end{equation}
		where $f_T(t)$ is given in \eqref{eq_section2_new1}. Replacing the respective PDFs in \eqref{eq_app_h1} and changing the order of integration and summation, yields
		
		\begin{equation}
f_R(r)\! = \!\!\sum _{i=0}^{\infty } \!\frac{8 \left[(m_d -1)\kappa \right]^{\frac{m_d+i}{2} } \mathcal{K}^{i+\mu } \mu ^{\frac{i+m_d}{2}} r^{2 i+2 \mu -1} 
			m_t^{m_t}}{\hat{r}^{2 i+2 \mu } i! \Gamma \left(m_d\right) \Gamma (i+\mu ) \Gamma \left(m_t\right)} \!{\rm{K}}_{m_d-i}\!\left(\!2 \sqrt{(m_d\!-\!1) \mu  \kappa }\right) \!\!\!\int_0^{\infty} \!\!\frac{ a^{2 \left(m_t-i-\mu \right)-1}}{ {\rm{e}}^{a^2 m_t+\frac{r^2 \mathcal{K}}{a^2 \hat{r}^2}}} \, da.
		\end{equation}
		Now, solving the above integral using~\cite[eq. 2.3.16.1]{prudinkov_v1} and after some algebraic manipulations, we obtain~\eqref{eq_section3_10}.
	}

	%%%%%%%%%%%%%%%%%%%%%%%%%%%%%%%%%%%%%%%%%%%%%%%%%%%%%%%%%%%
	\subsection{F. PROOF OF THEOREM~\ref{thm Type2pdf}$\label{app:F}$}
	%%%%%%%%%%%%%%%%%%%%%%%%%%%%%%%%%%%%%%%%%%%%%%%%%%%%%%%%%%%	
	We determine the envelope distribution, $R$, of the double shadowed $\kappa$-$\mu$ Type~II~(example~1) fading model when $A$ and $B$ vary according to the inverse Nakagami-$m$ and Nakagami-$m$ distributions, respectively, from the following integral
	
	\begin{equation}
	f_{R}\left(r\right)=\intop_{0}^{\infty}\intop_{0}^{\infty}f_{R|\alpha,\beta}\left(r|\alpha,\beta\right)f_{\alpha}\left(\alpha\right)f_{\beta}\left(\beta\right)d\alpha d\beta
	\label{eq_app_f1}
	\end{equation}
		where 
		\begin{equation}
	f_{R|\beta}\left(r|\beta\right)=\intop_{0}^{\infty}f_{R|\alpha,\beta}\left(r|\alpha,\beta\right)f_{\alpha}\left(\alpha\right)d\alpha 
	\label{eq_app_f2}
	\end{equation}
		and the double shadowed $\kappa$-$\mu$ Type~II~(example~1) signal model insinuates that $f_{R|\alpha,\beta}\left(r|\alpha,\beta\right)$ follows a $\kappa$-$\mu$ distribution with PDF~\cite{4231253} 
	
	\begin{equation}
	f_{R|\alpha,\beta}\left(r|\alpha,\beta\right)=\frac{r^{\mu}}{\sigma^{2}\alpha^{2}\left(\beta d\right)^{\mu-1}}\mathrm{e}^{\frac{-r^{2}-\left(\beta d\right)^{2}}{2\sigma^{2}\alpha^{2}}}\mathrm{I}_{\mu-1}\left(\frac{\beta dr}{\sigma^{2}\alpha^{2}}\right)
	\label{eq_app_f3}
	\end{equation}
		whilst $f_{\alpha}\left(\alpha\right)$ is similar to~\eqref{eq_app_a2} where $\xi$ and $m_d$ are replaced with $\alpha$ and $m_s$, respectively. Likewise, $f_{\beta}\left(\beta\right)$ is similar to~\eqref{eq_app_a1} where $\xi$ is replaced with $\beta$, and $m_s$ is replaced with $m_d$. Making appropriate substitutions in~\eqref{eq_app_a2}, followed by using the resultant expression and~\eqref{eq_app_f3} in~\eqref{eq_app_f2}, and finally solving the integral using~\cite[eq. 2.15.3.2]{prudinkov} we obtain
	
	\begin{equation}
	\!f_{R|\beta}\left(r|\beta\right)=\frac{2^{m_{s}+1}(m_{s}-1)^{m_{s}}r^{2\mu-1}\sigma^{2m_{s}}\Gamma\left(m_{s}+\mu\right){}_{2}F_{1}\!\!\left(\!\frac{m_{s}\!+\!\mu}{2};\theta_1;\mu;\frac{4d^{2}r^{2}\beta^{2}}{\left(r^{2}+d^{2}\beta^{2}+2(m_{s}-1)\sigma^{2}\right)^{2}}\!\right)}{\Gamma\left(m_{s}\right)\Gamma\left(\mu\right)\left(r^{2}+d^{2}\beta^{2}+2(m_{s}-1)\sigma^{2}\right)^{m_{s}+\mu}}.
	\label{eq_app_f4}
	\end{equation}
		Substituting~\eqref{eq_app_f4} and \eqref{eq_app_a1} (after replacing $\xi$ and $m_s$ with $\beta$ and $m_d$) in~\eqref{eq_app_f1}, simplifying the resultant integral, followed by replacing the Gauss hypergeometric function with its series representation~\cite[07.23.02.0001.01]{wolfram} we obtain
	
	\begin{align}
	&f_{R}\left(r\right)=\frac{2^{m_{s}+2}(m_{s}-1)^{m_{s}}r^{2\mu-1}\sigma^{2m_{s}}m_{d}^{m_{d}}\Gamma\left(m_{s}+\mu\right)}{\Gamma\left(m_{s}\right)\Gamma\left(m_{d}\right)\left(r^{2}+2(m_{s}-1)\sigma^{2}\right)^{m_{s}+\mu}} \nonumber \\ &\!\!\times \sum_{i=0}^{\infty}\frac{\left(\frac{m_{s}+\mu}{2}\right)_i\left(\frac{m_{s}+\mu+1}{2}\right)_i\left(4d^{2}r^{2}\right)^{i}}{i!\Gamma\left(\mu+i\right)\left(r^{2}+2(m_{s}-1)\sigma^{2}\right)^{2i}} \!\intop_{0}^{\infty} \!\frac{\beta^{2i+2m_{d}-1}}{\mathrm{e}^{m_{d}\beta^{2}}} \! \left(\frac{d^{2}\beta^{2}}{r^{2}+2(m_{s}-1)\sigma^{2}}+1\!\right)^{\!\!{-2i-\mu-m_{s}}} d\beta. 
	\label{eq_app_f5}
	\end{align}
		Now solving the integral in~\eqref{eq_app_f5} using~\cite[eq. 13.2.5]{Abramowitz1972}, followed by substituting~\cite[07.33.17.0007.01]{wolfram} for the hypergeometric $\mathrm{U}$ function (Tricomi confluent hypergeometric function), $d = \sqrt{2 \mu {\sigma^{2}} \kappa}$ and $\sigma=\sqrt{\frac{\hat{r}^{2}}{2\mu\left(1+\kappa\right)}}$, and finally simplifying the resultant expression we obtain~\eqref{eq_section3_08}.

%%%%%%%%%%%%%%%%%%%%%%%%%%%%%%%%%%%%%%%%%%%%%%%%%%%%%%%%%%%
{\subsection{G. PROOF OF THEOREM~\ref{thm TypeVIpdf}$\label{app:J}$}
	%%%%%%%%%%%%%%%%%%%%%%%%%%%%%%%%%%%%%%%%%%%%%%%%%%%%%%%%%%%	
	The envelope distribution, $R$, of the double shadowed $\kappa$-$\mu$ Type~II~(example~2) fading model when $A$ and $B$ vary according to Nakagami-$m$ and inverse Nakagami-$m$ distributions, respectively, can be obtained through \eqref{eq_app_f1}. The double shadowed $\kappa$-$\mu$ Type~II~(example~2) signal model presented in~\eqref{eq_section3_07} insinuates that $f_{R|\alpha,\beta}(r|\alpha,\beta)$ follows a $\kappa$-$\mu$ distribution with PDF given in~\eqref{eq_app_f3}. Also, $f_\alpha(\alpha)$ is similar to~\eqref{eq_app_a1} with $\xi$ replaced by $\alpha$, and $f_\beta(\beta)$ is similar to \eqref{eq_app_a2} with $\xi$ replaced by $\beta$. Now integrating with respect to $\beta$, we obtain an expression similar to~\eqref{eq_section2_new1} which is conditioned on $\alpha$
	
	\begin{align}
	&f_R(r) = \frac{8 \mathcal{K}^{\mu } \left(\kappa  \mu  (m_d-1)\right)^{\frac{m_d}{2}} m_s^{m_s} r^{2 \mu -1}}{\Gamma \left(m_d\right) \Gamma \left(m_s\right) \hat{r}^{2 \mu }} \sum _{i=0}^{\infty } \frac{1}{i! \Gamma (\mu +i)} \left(\frac{r^2 \mathcal{K} \sqrt{\kappa  \mu  (m_d-1)}}{\hat{r}^2}\right)^i \nonumber \\
	& \times \int_0^{\infty } \alpha ^{-1-3 i-2 \mu -m_d+2 m_s} e^{-\left(\alpha ^2 m_s+\frac{r^2 \mathcal{K}}{\alpha ^2 \hat{r}^2}\right)} {\rm{K}}_{i-m_d}\left(\frac{2 \sqrt{\kappa  \mu  (m_d-1)}}{\alpha }\right) d\alpha.
	\end{align} 
	The above integral can be solved by replacing the Bessel function with its power series representation \cite[03.04.06.0002.01]{wolfram} followed by changing the order of integration and summation. Now using~\cite[eq. 2.3.16.1]{prudinkov_v1} we obtain
	
\begin{align}
		f_R(r)& =  \frac{4 \pi  \mathcal{K}^{\mu } m_s^{m_s} r^{2 \mu-1 }\hat{r}^{-2 \mu }}{\sin \left(\pi  m_d\right) \Gamma \left(m_d\right) \Gamma \left(m_s\right) } \Bigg(\sum _{i=0}^{\infty } \sum _{k=0}^{\infty } \frac{(-1)^i m_s^i}{i! k! \Gamma (i+\mu )}  \frac{\left(\kappa  \mu  (m_d-1)\right)^{k+i}}{\Gamma \left(1+i+k-m_d\right)}\left(\frac{r^2 \mathcal{K}}{\hat{r}^2 m_s}\right)^{\frac{m_s-k-\mu}{2}} \nonumber \\ &\times {\rm{K}}_{2 i+k+\mu -m_s}\left(\frac{2 r \sqrt{\mathcal{K} m_s}}{\hat{r}}\right) 
	-\sum _{i=0}^{\infty } \sum _{k=0}^{\infty } \frac{(-1)^i m_s^i}{i! k! \Gamma (i+\mu )}\frac{\left(\kappa  \mu  (m_d-1)\right)^{k+m_d}}{\Gamma \left(1-i+k+m_d\right)} \left(\frac{r^2 \mathcal{K}}{\hat{r}^2 m_s}\right)^{\frac{i-k-\mu -m_d+m_s}{2} } \nonumber \\ &\times {\rm{K}}_{i+k+\mu +m_d-m_s}\left(\frac{2 r \sqrt{\mathcal{K} m_s}}{\hat{r}}\right)\Bigg) \label{eq_app_j2}
		\end{align}
	The first double summation in~\eqref{eq_app_j2} can be simplified by using the index substitution $2i+k=n$ so that the inner sum (after some algebraic manipulations) simplifies to 
	
	\begin{equation}
\sum _{i=0}^{n/2} \frac{\left(-\frac{n}{2}\right)_i \left(\frac{1-n}{2}\right)_i \left(-n+m_d\right)_i}{i! (\mu )_i\Gamma(\mu)} \!\left(\frac{4 r^2 (1+\kappa )}{\hat{r}^2 \kappa  (m_d-1)}\right)^i 
	\! =\! \, _3\tilde{F}_1\left(-\frac{n}{2},\frac{1\!-\!n}{2},m_d-n;\mu ;\frac{4 r^2 (1+\kappa )}{\hat{r}^2 \kappa  (m_d-1)}\right).
	\end{equation}
	Further simplifying and changing the index $n$ to $i$, we obtain the first part of~\eqref{eq_section3_14}. The last double summation in~\eqref{eq_app_j2} is simplified by summing over the infinite triangle $k=n-i$. The inner sum now reduces to 
	
	\begin{equation}
     \sum _{i=0}^n \frac{(-n)_i \left(\frac{-n-m_d}{2} \right)_i \left(\frac{1-n-m_d}{2}\right)_i}{i! (\mu )_i\Gamma (\mu )} \left(\frac{4 r^2 (1+\kappa )}{\hat{r}^2 \kappa  (m_d-1)}\right)^i
	 \!\! =\! \, _3\tilde{F}_1\!\left(\!-n,\frac{-n\!-\!m_d}{2} ,\frac{1\!-\!n\!-\!m_d}{2} ;\mu ;\frac{4 r^2 (1+\kappa )}{\hat{r}^2 \kappa  (m_d-1)}\!\right)\nonumber
	 \end{equation}
		 which after some algebraic manipulations simplifies to the second part of~\eqref{eq_section3_14}.}	
	
%%%%%%%%%%%%%%%%%%%%%%%%%%%%%%%%%%%%%%%%%%%%%%%%%%%%%%%%%%%
\subsection{H. PROOF OF THEOREM~\ref{thm TypeIVpdf}$\label{app:I}$}
	%%%%%%%%%%%%%%%%%%%%%%%%%%%%%%%%%%%%%%%%%%%%%%%%%%%%%%%%%%%	
		The double shadowed $\kappa$-$\mu$ Type~III~(example~1) model can be viewed as a product of a Nakagami-$m$ RV and a single shadowed $\kappa$-$\mu$ Type~II~(example~1) RV. Accordingly, its PDF can be obtained by first replacing~\eqref{eq_section2_06a} with the hypergeometric function expressed in terms of its power series expression~\cite[eq. 15.1.1]{Abramowitz1972}, then substituting the resultant expression and~\eqref{eq_app_a1} (after replacing $\xi$ and $m_s$ with $\alpha$ and $m_t$) in~\eqref{eq_app_h1}, changing the order of integration and summation, and finally followed by some algebraic manipulations as
		
	\begin{align}
		& f_R(r) = \frac{4 \left(m_s-1\right)^{m_s} m_t^{m_t} \hat{r}^{2 m_s}}{\mathcal{K}^{m_s} \Gamma \left(m_t\right) B\left(m_s,\mu \right) r^{2 m_s+1}} \sum _{i=0}^{\infty } \frac{\left(\frac{m_s+\mu}{2} \right)_i \left(\theta _1\right)_i}{i! (\mu )_i} 
	 \left(\frac{4 \kappa  \hat{r}^2}{r^2 (1+\kappa )}\right)^i \nonumber \\
		& \times\int_0^{\infty } \alpha ^{-1+2 i+2 m_s+2 m_t}   \left(1+\frac{\alpha ^2 \hat{r}^2 \left(m_s-1+\kappa  \mu \right)}{r^2 (1+\kappa ) \mu }\right)^{-2 i-\mu -m_s} {\rm{e}}^{-m_t \alpha ^2} d\alpha.
		\end{align}
Solving the above integral using~\cite[eq. 13.2.5]{Abramowitz1972} followed by some algebraic manipulations yields~\eqref{eq_section3_12}.

%%%%%%%%%%%%%%%%%%%%%%%%%%%%%%%%%%%%%%%%%%%%%%%%%%%%%%%%%%%
	\subsection{I. PROOF OF THEOREM~\ref{thm TypeVpdf}$\label{app:K}$}
%%%%%%%%%%%%%%%%%%%%%%%%%%%%%%%%%%%%%%%%%%%%%%%%%%%%%%%%%%%	
The double shadowed $\kappa$-$\mu$ Type~III~(example~2) model can be viewed as a single shadowed $\kappa$-$\mu$ Type~III~(example~2) model in which the variation of the scattered waves is influenced by a Nakagami-$m$ RV. The PDF of the envelope $R$ can be obtained via 

\begin{equation}
f_R(r) = \int_0^\infty f_{R|\alpha}(r|\alpha) f_\alpha(\alpha)\, d\alpha \label{eq_app_I1}
\end{equation}
where $f_\alpha(\alpha)$ is similar to \eqref{eq_app_a1} such that $\xi$ is replaced by $\alpha$. $f_{R|\alpha}(r|\alpha)$ can be obtained from~\eqref{eq_section2_08} by first replacing $\kappa = d^2/(2\mu \sigma^2)$ and $\hat{r} = 2\mu \sigma^2+d^2$, then multiplying $\sigma$ by $\alpha$ and finally using $d=\sqrt{2\mu\sigma^2\kappa}$; $\sigma=\sqrt{\frac{\hat{r}^2}{2\mu(1+\kappa)}}$ as follows:

\begin{align}
f_{R|\alpha}(r|\alpha)& = \frac{2  (m_t-1)^{m_t}\hat{r}^{2 m_t} \alpha ^{2 m_t}}{((1+\kappa ) \mu )^{m_t}B\left(m_t,\mu \right) r^{1+2 m_t}}   
\left(1+\frac{\alpha ^2 \hat{r}^2 (m_t-1)}{r^2 (1+\kappa ) \mu }\right)^{-\mu -m_t}  \nonumber \\
&\times \exp \left(-\frac{\kappa  \mu }{\alpha ^2}\right) \, _1F_1\left(\mu +m_t;\mu ;\frac{\kappa  \mu }{\alpha ^2 }\left(1+\frac{\alpha ^2 \hat{r}^2 (m_t-1)}{r^2 (1+\kappa ) \mu }\right)^{-1}\right).
 \label{eq_app_I2}
\end{align}
Now substituting \eqref{eq_app_I2} in \eqref{eq_app_I1} we obtain

\begin{align}
f_R(r) &= \frac{4 ((1+\kappa ) \mu )^{-m_t} (m_t-1)^{m_t} m_s^{m_s} r^{-1-2 m_t}}{B\left(m_t,\mu \right) \Gamma \left(m_s\right) \hat{r}^{-2 m_t}} \int_0^{\infty } \frac{\alpha ^{-1+2 m_t+2 m_s}}{\exp \left(\alpha ^2 m_s+\frac{\kappa  \mu }{\alpha ^2}\right)} \nonumber \\
&\times \left(1+\frac{\alpha ^2 \hat{r}^2 (m_t-1)}{r^2 (1+\kappa ) \mu }\right)^{-\mu -m_t} \, _1F_1\left(\mu +m_t;\mu ;\frac{\kappa  \mu }{\alpha ^2}\left(1+\frac{\alpha ^2 \hat{r}^2 m_t}{r^2 (1+\kappa ) \mu }\right)^{-1}\right) \, d\alpha. 
\end{align}
It is possible to rewrite the hypergeometric function in terms of its Mellin-Barnes contour integral representation using~\cite[eq. 7.2.3.12]{prudinkov_v3}, whilst the exponential function can be written as a product of two contour integrals using \cite[eq. 8.4.3.1 and eq. 8.4.3.2]{prudinkov_v3} and \cite[eq. 8.2.1.1]{prudinkov_v3}. Now performing some algebraic manipulations we obtain

\begin{align}
&f_R(r)=\frac{4 \mathcal{K}^{-m_t} (m_t-1)^{m_t}  \hat{r}^{2 m_t} }{\Gamma \left(m_t\right) \Gamma \left(m_s\right)r^{1+2 m_t}} \left(\frac{1}{2 \pi  j}\right)^3 \int_0^\infty \!\!\oint_\mathcal{L} \Gamma \left(t_1\right)  \frac{\Gamma \left(-t_2\right) \Gamma \left(-t_3\right) \Gamma \left(t_3+\mu +m_t\right)(\kappa  \mu )^{t_2+t_3}}{\Gamma \left(t_3+\mu \right)(-1)^{t_3} m_s^{t_1-m_s}}  \nonumber \\
& \times \alpha ^{2 \theta_3-1} \left(1+\frac{\alpha ^2 \hat{r}^2 (m_t-1)}{r^2 \mathcal{K} }\right)^{-t_3-\mu -m_t} dt_1 dt_2 dt_3 d\alpha
\end{align}
where $\theta_3 = m_t+m_s-t_1-t_2-t_3$, $\mathcal{K}$ is as defined previously, $j = \sqrt{-1}$ is the imaginary particle and $\mathcal{L}$ is a suitable contour in the complex space. Now changing the order of integration, the inner integral can be solved using \cite[eq. 2.2.5.24]{prudinkov_v1}, which results in the triple contour integral as follows:

\begin{align}
f_R(r)& = \frac{2 (m_s\mathcal{K})^{m_s}   r^{2 m_s-1}}{\Gamma \left(m_t\right) \Gamma \left(m_s\right)(m_t-1)^{m_s} \hat{r}^{2 m_s} } \left(\frac{1}{2 \pi  j}\right)^3 \!\!\!\!  \oint_\mathcal{L} \Theta(t_1,t_2,t_3)\left(\frac{r^2 \mathcal{K} m_s}{\hat{r}^2 (m_t-1)}\right)^{\!\!\!{-t_1}}\nonumber \\ &\times \left(\frac{r^2 \mathcal{K}}{\kappa  \mu  \hat{r}^2 (m_t\!-\!1)}\right)^{\!\!\!{-t_2}}\left(\frac{-r^2 \mathcal{K}}{\kappa  \mu  \hat{r}^2	(m_t\!-\!1)}\right)^{-t_3}dt_1dt_2dt_3
\end{align}
where
\begin{equation}
\Theta(t_1,t_2,t_3) = \frac{\Gamma \left(t_1\right) \Gamma \left(-t_2\right) \Gamma \left(-t_3\right) \Gamma \left(\theta_3\right) \Gamma \left(\theta _4\right)}{\Gamma \left(t_3+\mu \right)}
\end{equation}
and $\theta_4 = t_1+t_2+2 t_3+\mu -m_s$. It is possible to obtain a multi-fold series representation from the above contour integral through the sum of residues theorem. The residues for the variable $t_1$ are taken around the poles of $\Gamma(t_1)$ and $\Gamma(\theta_4)$; the residues for $t_2$ are taken from $\Gamma(-t_2)$ and $\Gamma(\theta_3)$; and the residues for $t_3$ are taken from $\Gamma(-t_3)$. This results in 
  %at the top of the next page.
\begin{equation}
\begin{split}
&f_R(r) =\frac{2 \left(\mathcal{K} m_s/(m_t-1)\right)^{m_s} r^{2 m_s-1}}{\Gamma \left(m_t\right) \Gamma \left(m_s\right) \hat{r}^{2 m_s}}(S_1 + S_2 + S_3)
\end{split}
\end{equation}
where 
		\begin{equation}
	\!\!\!\!\!\!\!\!S_1 = \!\!\!\!\!\sum_{i,j,k'=0}^\infty\!\!\!\frac{(-1)^{i+j} \Gamma \left(-i\!+\!j\!+\!2 k'\!+\!\mu\! -\!m_s\right) \Gamma \left(i\!-\!j\!-\!k'\!+\!m_t\!+\!m_s\right)}{i! j! k'! \Gamma (k'\!+\!\mu )}\! \left(\frac{r^2 \mathcal{K} m_s}{\hat{r}^2 (m_t\!-\!1)}\right)^{\!\!i} \! \left(\frac{\kappa  \mu  \hat{r}^2 (m_t\!-\!1)}{r^2 \mathcal{K}}\right)^{\!\!{j+k'}}\label{eq_app_I3}
	\end{equation}
	\begin{equation}
	\!\!\!\!\!\!\!\!S_2 = \!\!\!\!\!\sum_{i,j,k'=0}^\infty\!\!\frac{(-1)^{i+j} \Gamma \left(j\!+\!k'\!+\!\mu\! +\!m_t\right) \Gamma \left(-i\!-\!j\!+\!k'\!-\!m_t\!-\!m_s\right)}{i! j! k'! \Gamma (k'\!+\!\mu )} \left(\kappa  \mu  m_s\right)^{\!i} \left(\frac{\kappa  \hat{r}^2 (m_t\!-\!1)}{r^2 (1\!+\!\kappa )}\right)^{\!\!{j+m_t+m_s}} \label{eq_app_I5}
	\end{equation}
	\begin{equation}
	\!\!\!\!\!\!\!\!S_3 = \!\!\!\!\!\sum_{i,j,k'=0}^\infty\!\!\!\frac{(-1)^{i+j} \Gamma \left(i+k'+\mu +m_t\right) \Gamma \left(-i\!-\!j\!-\!2 k'\!-\!\mu\! +\!m_s\right)}{i! j! k'! \Gamma (k'\!+\!\mu )\left(\kappa  \mu  m_s\right)^{\!{-j}}} \!\left(\!\frac{r^2 \mathcal{K} m_s}{\hat{r}^2 (m_t\!-\!1)}\right)^{\!\!{i+2 k'+\mu-m_s}} \!\! \left(\!\frac{\kappa  \mu  \hat{r}^2 (m_t\!-\!1)}{r^2 \mathcal{K}}\right)^{\!\!{k'}} \label{eq_app_I4}
	\end{equation}
The first triple summation $S_1$ can be reduced by summing over the infinite triangle $j=n-k$ and using~\cite[eq. 4.2.5.55]{prudinkov_v1}, yielding 

\begin{equation}
S_1\! =\! \sum_{i,n=0}^\infty \frac{\pi  \csc \left(\pi  \left(\mu\! -\!m_s\right)\right)  \Gamma \left(i\!-\!n\!+\!m_t\!+\!m_s\right)}{i! n! \Gamma (n\!+\!\mu )  \Gamma \left(1\!+\!i\!-\!n\!-\!\mu\! +\!m_s\right)}  \frac{\Gamma \left(i\!+\!m_s\right)}{\Gamma \left(i\!-\!n\!+\!m_s\right)} \left(\frac{\kappa \mu \hat{r}^2 (m_t\!-\!1)}{r^2 \mathcal{K}}\right)^{\!n} \left(\frac{r^2 \mathcal{K}  m_s}{\hat{r}^2 (m_t\!-\!1)}\right)^{\!i}.
\label{eq_app_S1}
\end{equation}
The triple summation $S_2$ can be simplified by summing it over index $k'$. This results in a Gauss hypergeometric function whose argument is one. Using \cite[eq. 7.3.5.2]{prudinkov_v3}, we obtain

\begin{equation}
S_2 = \sum_{i,j=0}^\infty \frac{(-1)^{i+j}  \Gamma \left(-i-j-m_t-m_s\right) \Gamma \left(i\!+\!m_s\right)}{i! j!  \Gamma \left(i\!+\!j\!+\!\mu\! +\!m_t\!+\!m_s\right)}\frac{\Gamma \left(j\!+\!\mu\! +\!m_t\right)}{\Gamma \left(-j\!-\!m_t\right)} \!\left(\frac{\kappa \mu \hat{r}^2 (m_t\!-\!1)}{r^2 \mathcal{K}}\right)^{j+m_t+m_s}\!\!\!\left(\kappa  \mu  m_s\right)^i.
\label{eq_app_S2}
\end{equation}
To reduce $S_3$, it is required to first perform the variable transformation $j=n-k'$ followed by $i = j-k'$, then performing some algebraic manipulations the inner sum on the index $k'$ is solved using~\cite[eq. 4.2.5.25 ]{prudinkov} as follows:

\begin{equation}
\sum _{k'=0}^{n} \!\!\frac{1}{(j\!-\!k')! k'! (n\!-\!k')! \Gamma (k'\!+\!\mu )}\!= \sum _{k'=0}^{n} \frac{\binom{n}{k'} \binom{j+\mu -1}{k'+\mu -1}}{n! \Gamma (j+\mu )} = \frac{\Gamma (j+n+\mu )}{j! n! \Gamma (j+\mu ) \Gamma (n+\mu )}.
\end{equation}
Now performing some algebraic manipulations, we obtain

\begin{equation}
S_3 = \sum_{j,n=0}^\infty \frac{(-1)^{j+n} \Gamma (j+n+\mu ) \Gamma \left(j+\mu +m_t\right) }{j! n! \Gamma (j+\mu ) \Gamma (n+\mu )} \Gamma \left(m_s-\mu-j-n\right) \left(\kappa  \mu  m_s\right)^n \left(\frac{r^2 \mathcal{K}  m_s}{\hat{r}^2 (m_t-1)}\right)^{j+\mu -m_s}.
\label{eq_app_S3}
\end{equation}
Now summing~\eqref{eq_app_S1} over the index $n$, \eqref{eq_app_S2} over index $j$ and \eqref{eq_app_S3} over the infinite triangle $j=i-n$, followed by some algebraic manipulations the double shadowed $\kappa$-$\mu$ Type~III~(example~2) PDF simplifies to \eqref{eq_section3_13}, which completes the proof.

\bibliographystyle{IEEEtran}
\bibliography{IEEEabrv,ref}

\end{document}